\documentclass[a4paper,11pt]{article}
\pdfoutput=0 

\usepackage{jcappub} 
\usepackage{graphicx}
\usepackage{subfigure}
\usepackage[T1]{fontenc} 
\usepackage{tikz}

\usepackage{amssymb, amsmath, epsfig, natbib}

\renewcommand{\eqref}[1] {equation $($\ref{#1}$)$}
\newcommand{\comment}[1]{{}}
\def\beq{\begin{equation}}
\def\eeq{\end{equation}}
\def\beqn{\begin{eqnarray}}
\def\eeqn{\end{eqnarray}}

\def\2gcm{\textrm{g cm$^{-2}$}}

\def\H0{\ensuremath{\mathrm{H}_0}}

\newcommand{\bx}{{\bf x}}
\newcommand{\bk}{{\bf k}}
\newcommand{\bq}{{\bf q}}
\newcommand{\bP}{{\bf\Psi}}
\newcommand{\bs}{{\bf s}}

\author[a,b]{Blake D.~Sherwin,}
\author[c,d,e]{Martin White}

\affiliation[a]{Department of Applied Mathematics and Theoretical Physics, University of Cambridge,
Cambridge CB3 0WA, UK}
\affiliation[b]{Kavli Institute for Cosmology Cambridge, University of Cambridge,
Cambridge CB3 0HA, UK}
\affiliation[c]{Physics Division, Lawrence Berkeley National Laboratory,
Berkeley, CA 94720, USA}
\affiliation[d]{Department of Physics, University of California,
Berkeley, CA 94720}
\affiliation[e]{Department of Astronomy, University of California,
Berkeley, CA 94720}

\emailAdd{sherwin@damtp.cam.ac.uk}
\emailAdd{mwhite@berkeley.edu}

\title{The Impact of Wrong Assumptions in BAO Reconstruction}


\abstract{The process of density field reconstruction enhances the statistical power of distance scale measurements using baryon acoustic oscillations (BAO).  During this process a fiducial cosmology is assumed in
order to convert sky coordinates and redshifts into distances; fiducial bias and redshift-space-distortion parameters are also assumed in this procedure. We analytically assess the impact of incorrect cosmology and bias assumptions on the post-reconstruction power spectra using low-order Lagrangian perturbation theory, deriving general expressions for the incorrectly reconstructed spectra. We find that the BAO peak location appears to shift only by a negligible amount due to wrong assumptions made during reconstruction. However, the shape of the BAO peak and the quadrupole amplitude can be affected by such errors (at the percent- and five-percent-level respectively), which potentially could cause small biases in parameter inference for future surveys; we outline solutions to such complications.}

\arxivnumber{18MM.NNNNN}

\begin{document}
\maketitle
\flushbottom

\section{Introduction}

The measurement of baryon acoustic oscillations in the clustering of
galaxies and the intergalactic medium provides some of our tightest
constraints on the cosmological distance scale
(for a review of recent measurements see \cite{BOSS_DR12}, e.g.~Fig.~14,
 for reviews of the methods see \cite{Wei13,PDG14}).
With the exception of the intergalactic medium results, all recent measurements have employed a process known as ``reconstruction'' \cite{ESSS07}, which aims to undo some of the loss of signal due to non-linear evolution by estimating the motion of tracers under gravity and reversing it.
In order to perform reconstruction, one typically assumes a fiducial
cosmology to convert sky positions and redshifts into comoving coordinates; one also assumes a fiducial bias and redshift-space distortion (RSD) parameter to derive the density field from the galaxy distribution.
However, the cosmology assumed may not match the true, underlying cosmology, and the fiducial bias and RSD parameters may also not equal their true values. In this paper we use Lagrangian perturbation theory to assess the impact of such errors in the assumed cosmology, biasing and RSD upon the two-point clustering of the reconstructed field.

Reconstruction has been extensively studied in the literature using both
N-body simulations and analytic models
\cite{ESSS07,PWC09,NWP09,Seo10,Pad12,TasZal12b,McCSza12,SheZal12,
      Xu13,Sch15,BPH15,AchBla15,Whi15a,Coh16,Seo16,Var16,Hik17}.
While the impact of incorrect choices of distance, growth rate and bias has been studied in numerical simulations in particular cases, we are not aware of any analytic treatment.
Since the standard algorithm is based upon the Zel'dovich approximation
\cite{Zel70}, we choose to similarly use low-order Lagrangian perturbation theory
to assess the impact of incorrect assumptions about the cosmological model
upon the recovered statistics.  While the results will only be approximate,
due to our approximate model of structure formation, they allow us to
provide general formulae and hence estimate the errors introduced in any scenario.

The outline of the paper is as follows.
In Section \ref{sec:background} we briefly review Lagrangian perturbation theory and reconstruction and define our notation.
Section \ref{sec:distances_wrong} explains the calculation of the reconstructed power spectra in a simple case where only the assumed distances are incorrect. Section \ref{sec:everything_wrong} describes the more general calculation where the distances, the growth rate and the bias assumed can all be wrong.
In Section \ref{sec:discussion}, we evaluate our calculations and discuss the results. We conclude in Section \ref{sec:conclusions}.

\section{Background}
\label{sec:background}

This section reviews material which has been extensively discussed in the
literature already and is included primarily to set notation and present
some results which will be of use later.

We choose to study reconstruction within the framework of Lagrangian
perturbation theory (largely following Refs.~\cite{Mat08a,Mat08b}).
Lagrangian theory describes the evolution of large scale structure as
a mapping from the original positions of a fluid element, $\bq$, to the
final position, $\bx$, via a displacement vector, $\bP$.
Mass conservation then implies that the Fourier-space mass density in Eulerian
coordinates can be written
\beq
  \delta(\bk) =  \int d^3q\ e^{-i\bk\cdot\bq}
  \left( e^{-i \bk \cdot \bP(\bq)} - 1 \right).
\label{eqn:delta_defn}
\eeq
As a first step we will neglect bias and redshift-space distortions.
The computation of the power spectrum, at lowest order in (resummed) Lagrangian
perturbation theory, then follows a standard procedure \cite{Mat08a,Mat08b}.
Let us quickly review the steps. We compute $\langle\delta(\bk_1)\delta^\star(\bk_2)\rangle$ and use the cumulant
theorem to evaluate the expectation value of an exponential:
\beq
  \int d^3 q e^{-\bk \cdot \bq}
  \left(\left\langle
  e^{-i\left[\bk\cdot\bP(\bq_1)-\bk\cdot\bP(\bq_2)\right]}
  \right\rangle-1\right)
  \simeq \int d^3 q e^{-\bk \cdot \bq}  \left(e^{-k^2\Sigma^2/2}\exp\left[2\,k_i k_j \xi_{ij}(q)\right]-1\right),
\label{eqn:cumulant}
\eeq
where $q=\left|\bq_1-\bq_2\right|$,
$\xi_{ij}(q)=\left\langle\Psi_i(\bq_1)\Psi_j(\bq_2)\right\rangle$ is the
Lagrangian 2-point function and we have defined
\beq
  \Sigma^2 = \int \frac{dk}{3\pi^2}\ P_L(k).
\label{eqn:Sigma2}
\eeq
We caution the reader that there are other definitions of $\Sigma^2$ in the literature,
differing by factors of 2.
Expanding\footnote{Expanding the non-zero-lag piece of the correlator from
the exponential while keeping the zero-lag ($\Sigma^2$) terms exponentiated
is a reasonable approximation on large scales for CDM cosmologies, but can be
problematic in general.  See discussion in Refs.~\cite{TasZal12a,CLPT,Whi15a}.}
the second, non-zero-lag (i.e., $\bq_1 \neq \bq_2$) exponential in Eq.~(\ref{eqn:cumulant}) to linear order in the $\xi_{ij}$
and dropping a constant term, the Fourier transform is simply the linear power spectrum
$P_L(k)$. We thus obtain
$P(k) = e^{-k^2 \Sigma^2/2}\ P_L(k) + \cdots$. 
This result reveals the well known fact that features in the power spectrum,
such as the baryon acoustic oscillations, are damped by non-linear structure
formation\footnote{In configuration space, the peak in the correlation function
is convolved with a Gaussian of width $\Sigma$ and thus broadened.}
\cite{Bha96}.
Had we kept higher order terms (the $\cdots$), they would have included mode
coupling terms which give a slight shift to the oscillations in the power
spectrum or the location of the peak in the correlation function.  These
mode coupling terms are small, and they will be further multiplied by small factors of order the fractional errors in cosmology, bias and RSD; we will therefore neglect them in what follows.

Continuing to neglect bias and redshift-space distortions, the standard
reconstruction algorithm \cite{ESSS07} works as follows: we smooth the density field
with a filter, $S(k)$, to remove small scale modes and compute the negative
Zel'dovich displacement
\beq
  \bs (\bk) \equiv -i \frac{\bk}{k^2}\ S(k) \delta(\bk)
\eeq
The typical choice for the filter function, $S(k)$, is a Gaussian.
We then displace the original density field tracers by this vector to give
the ``displaced field''
\beq
  \delta_d(\bk) = \int d^3q\ e^{-i\bk \cdot \bq}
  \left( e^{-i\bk \cdot \left[ \bP(\bq)+\bs(\bq)\right]}-1 \right).
\eeq
We also displace a random set of particles by the same amount to give
the ``shifted field'':
\beq
  \delta_s(\bk) = \int d^3q\ e^{-i\bk \cdot \bq}
  \left( e^{-i\bk \cdot \left[\bs(\bq)\right]}-1 \right)
\eeq
and define the reconstructed density field as
$\delta_\mathrm{rec}(\bk) = \delta_d(\bk)-\delta_s(\bk)$.
It then follows that
\beq
  P_{\rm rec} = P_{dd}+P_{ss}-2P_{ds}.
\eeq
The computation of the reconstructed power spectrum $P_{\rm rec}$, at lowest order in
(resummed) Lagrangian perturbation theory, then follows the same steps as
above \cite{PWC09,NWP09,TasZal12b,McCSza12,SheZal12,Sch15,Whi15a,Coh16}.
Working to lowest order\footnote{Continuing our neglect of higher order terms, in what follows we shall neglect the differences between Eulerian and Lagrangian positions in defining $\mathbf{s}$, since this will be higher order in $\mathbf{\Psi}$, and assume that after smoothing $\delta$ can be replaced by $\delta_L$.  See Ref.~\cite{Whi15a} for further discussion.} the negative Zel'dovich displacement becomes $\bs(\bk) = -S(k) \bP(\bk)$ so the power spectrum of the shifted field is
\beq
  P_{ss}(k) = e^{-k^2 \Sigma_{ss}^2/2}\ S^2(k)\,P_L(k) +\cdots
  \label{simple1}
\eeq
where $\Sigma_{ss}^2$ is as for Eq.~(\ref{eqn:Sigma2}) but with $P_L$ multiplied
by $S^2$.  For the displaced field, the combination $\bP+\bs$ becomes
$[1-S(k)]\bP(\bk)$ so we obtain
\beq
  P_{dd}=e^{-k^2\Sigma_{dd}^2/2}\ [1-S(k)]^2\, P_L(k) + \cdots
\eeq
where $\Sigma_{dd}^2$ is as for Eq.~(\ref{eqn:Sigma2}) but with $P_L$ multiplied
by $[1-S]^2$. The cross-spectrum is
\beq
  P_{ds}=-e^{-k^2 \Sigma_{ds}^2/2} S(k)[1-S(k)]\, P_L(k) + \cdots
    \label{simple2}
\eeq
with $\Sigma^2_{ds}=(\Sigma^2_{ss}+\Sigma^2_{dd})/2$.

With these preliminaries in hand, we can now include bias and redshift-space
distortions.
For tracers that are locally biased in Lagrangian space, which we will
assume throughout, the integral in Eq.~(\ref{eqn:delta_defn}) is modulated
by a functional of the linear density field at the Lagrangian position:
$F[\delta_L(\bq)]$ \cite{Mat08b}.
We will need only the lowest moments of this functional, as we will work
throughout in the limit of linear bias which is appropriate on large scales.
Taking the Fourier transform of $F$ we can write \cite{Mat08a,Mat08b}
\beq
\delta(\bk) = \int \frac{d\lambda}{2\pi} d^3q\ F(\lambda) e^{-i\bk\cdot\bq+i \lambda\delta(\bq)} e^{- i\bk\cdot\bP(\bq)}
\qquad\quad (\bk\ne 0)
\eeq
Evaluating the exponential with the cumulant theorem and expanding to second order (though we again keep the zero-lag parts exponentiated) we obtain \cite{Mat08b}
\beqn
  P(k) = e^{-k^2 \Sigma^2/2}\left[ 1 + \langle F' \rangle\right]^2 P_L(k)
  + \cdots = e^{-k^2 \Sigma^2/2} b^2 P_L(k) + \cdots,
\eeqn
where we have defined $b=1+\left\langle F'\right\rangle$.

Redshift space distortions can also be added in a simple way in the Lagrangian perturbation theory formalism.
In Eq.~(\ref{eqn:delta_defn}) we change $\bP(\bq)$ to $\mathbf{R}\bP(\bq)$,
where the matrix $\mathbf{R}$ is defined via the RSD parameter $f$ as
$R_{ij}=(\delta_{ij}+f\hat{z}_i\hat{z}_j)$ if the line-of-sight\footnote{We
shall assume the plane-parallel approximation throughout.  For a discussion
of beyond plane-parallel terms in the Zel'dovich approximation see
Ref.~\cite{CasWhi18}.} is the $\hat{z}$ direction.
Using $\bk\cdot\mathbf{R}\bk=k^2(1+f\mu^2)$, with $\mu$ the cosine of the $\bk$-vector to the line of sight, the redshift-space power spectrum then follows directly as:
\beqn
 \frac{[\bk\cdot\mathbf{Rk}]^2}{k^4}P_L(k) \times \exp{\left[- k_\bot^2 \Sigma^2/2 - (1+f)^2 k_\parallel^2 \Sigma^2/2 \right]} \\=  [1+f \mu^2 ]^2 P_L(k) \times \exp{\left[- k_\bot^2 \Sigma^2/2 - (1+f)^2 k_\parallel^2 \Sigma^2/2 \right]}.
\eeqn
If we include bias at the same time the prefactor becomes $(b+f\mu^2)^2$ while the damping is unchanged.

There are different choices that one can make in implementing reconstruction
in the presence of redshift space distortions, and these are discussed in
Refs.~\cite{Whi15a,Seo16,Coh16}.  We follow the scheme in Ref.~\cite{Coh16}, again neglecting biasing at first.
The shift field is computed as
\beq
\mathbf{s}(\bk) = -\frac{i \bk}{k^2} S(k) \frac{\delta^{\rm obs}(k)}{1+f\mu^2} \simeq -\frac{i \bk}{k^2} S(k) \delta(k).
\eeq
The displaced and shifted densities are generated by $\mathbf{R}(\bP+\bs)$ and $\mathbf{R}\bs$ respectively.
Propagating these through a reconstruction analysis as discussed previously, we obtain
\beq
  P_{\rm rec}(\bk) = [1+f \mu^2 ]^2 P_L(k) D(\bk)
\eeq
where
\begin{align}
D(\bk) &= S^2(k)\exp\left[-\frac{1}{2}k_\bot^2\Sigma_{ss}^2
 - \frac{1}{2}(1+f)^2k_\parallel^2\Sigma_{ss}^2 \right] \nonumber \\
 &+ \left[1-S(k)\right]^2 \exp\left[-\frac{1}{2}k_\bot^2\Sigma_{dd}^2
 - \frac{1}{2}(1+f)^2k_\parallel^2\Sigma_{dd}^2\right] \nonumber \\
 &+ 2S(k)\left[1-S(k)\right] \exp\left[-\frac{1}{2}k_\bot^2\Sigma_{ds}^2
 - \frac{1}{2}(1+f)^2k_\parallel^2\Sigma_{ds}^2\right]
\end{align}
and the damping coefficients $\Sigma$ are defined as previously.
If we now add bias, it does not affect the damping factors, but the non-zero-lag terms differ, resulting in:
\beq
  P_{\rm rec}(\bk) = [b+f \mu^2 ]^2 P_L(k) D(\bk),
\eeq
with
\begin{align}
D(\bk) &= \left[\frac{1+f\mu^2}{b+f\mu^2}S(k)\right]^2
  \exp\left[-\frac{1}{2}k_\bot^2 \Sigma_{ss}^2
  -\frac{1}{2}(1+f)^2 k_\parallel^2 \Sigma_{ss}^2\right] \nonumber \\
  &+ \left(1-\left[\frac{1+f\mu^2}{b+f\mu^2}S(k)\right]\right)^2
  \exp\left[-\frac{1}{2}k_\bot^2 \Sigma_{dd}^2
  - \frac{1}{2}(1+f)^2 k_\parallel^2 \Sigma_{dd}^2\right] \nonumber \\
  &+ 2\left(\frac{1+f\mu^2}{b+f\mu^2}S(k)\right)
  \left(1-\left[\frac{1+f\mu^2}{b+f\mu^2}S(k)\right]\right)
  \exp\left[-\frac{1}{2}k_\bot^2 \Sigma_{ds}^2
  - \frac{1}{2}(1+f)^2 k_\parallel^2 \Sigma_{ds}^2\right],
\end{align}
which agrees with the expression in Ref.~\cite{Coh16}.

\section{Reconstruction with the wrong distances}
\label{sec:distances_wrong}

We now turn to the main focus of this paper, the impact of incorrect assumptions about the distance scale and dynamics when performing reconstruction. The impact of incorrect reconstruction assumptions has been studied in N-body
simulations (e.g.~\cite{Var16}) but has not been investigated analytically
before. The analytic theory turns out to be a relatively simple generalization of
results which have already appeared in the literature and can be used to
gain more insight into the effects that arise. 

In this section, we will begin by discussing the impact on reconstruction of inaccuracies in the distances. We simplify our treatment at first by considering the problem in real space, for the matter fields. In the next section we will generalize our analysis to include biasing and redshift space distortions, discussing reconstruction with incorrect fiducial growth rate and bias parameters.

We assume that the impact of an incorrect choice of cosmology can be
parameterized by a remapping the distance between pairs of objects as
\beq
  r^f_\bot= \alpha_\bot r_\bot' \qquad , \qquad
  r^f_\parallel= \alpha_\parallel r_\parallel'
\label{eqn:rescale}
\eeq
where $f$ indicates incorrectly assumed fiducial (or false) distances or other parameters, primes indicate true, physical distances, and we allow separate scalings transverse ($\alpha_\bot$) and parallel ($\alpha_\parallel$) to the line of sight.
We assume further that the region being analyzed is small enough that a constant scaling
is a good approximation over the relevant volume. What does this imply for our inferred, reconstructed power spectra?

We first note that the volume change induced by Eq.~(\ref{eqn:rescale}) does not
change our overdensities, since the effect on $\rho$ is canceled by the
effect on $\bar{\rho}$.  Hence we measure the following ``false'' density field:
\beq
  \delta^f(\bx)=\delta(\mathbf{A}^{-1}\bx)
  \quad {\rm with}\quad
  \mathbf{A}=\left( \begin{array}{ccc}
     \alpha_\perp & 0 & 0 \\
    0 & \alpha_\perp & 0\\
    0 & 0 & \alpha_\parallel \end{array} \right) \qquad .
\eeq
Here $\bx$ shall always refer to the ``observed'', false coordinates, i.e., we omit the superscript $\bx^f$. It will be helpful to recall that since $\mathbf{A}$ (and its inverse) are diagonal,
$(\mathbf{A}\mathbf{a})\cdot\mathbf{b}=(\mathbf{A}\mathbf{b})\cdot\mathbf{a}$
for any $\mathbf{a}$, $\mathbf{b}$.  

Fourier transforming $\delta^f(\bx)$ we obtain:
\beq
  \delta^f(\bk) = \alpha_\perp^2 \alpha_\parallel \delta(\mathbf{A} \bk)
  = \alpha_\perp^2 \alpha_\parallel
  \int d^3q\ e^{-i \mathbf{A} \bk \cdot \bq}
  \left(e^{-i\mathbf{A} \bk \cdot \bP(\bq)} - 1\right)
\eeq
where $\bk$, again omitting the superscript $^f$, is the conjugate wavenumber of the false coordinates.
The power spectrum is given by
\beq
  P^f(\bk) = \alpha_\perp^2 \alpha_\parallel\ P(\mathbf{A}\bk),
\eeq
recalling that the momentum conserving $\delta$-function has units of
volume.

We turn now to reconstruction. Substituting $\tilde\bq=\mathbf{A}\bq$ and then relabeling
$\tilde\bq\rightarrow\bq$ gives
\beq
  \delta^f(\bk) = \int d^3q\ e^{-i \bk \cdot \bq}
  \left( e^{-i \bk \cdot \left[ \mathbf{A} \bP(\mathbf{A}^{-1}  \bq) \right]}
  -1\right).
\eeq
The negative Zel'dovich displacement is
\beq
  \mathbf{s}^f(\bk) = -i\,\alpha_\perp^2 \alpha_\parallel
  \ \frac{\bk}{k^2}\ S(k)\,\delta(\mathbf{A}\bk)
\label{eqn:vecS_defn}
\eeq
If we define
\beq
  \mathbf{S}(\mathbf{A} \bk ) \equiv
  \left[ \frac{|\mathbf{A}\bk|^2S(k)}{k^2}\mathbf{A}^{-1}\mathbf{A}^{-1}\right],
\eeq
then using
\begin{equation}
  \mathbf{\Psi}(\mathbf{A}\mathbf{k}) = i\frac{\mathbf{Ak}}{|\mathbf{Ak}|^2}
  \delta(\mathbf{Ak})
\end{equation}
we can write
\beq
  \mathbf{s}^f(\bk) = -\alpha_\perp^2\alpha_\parallel
  \ \mathbf{S}(\mathbf{A}\bk)\,\mathbf{A}\bP(\mathbf{A}\bk)
  \quad\Rightarrow\quad
  \mathbf{s}^f(\bq)=-\mathbf{A} \mathbf{\tilde S}_{*}\bP(\mathbf{A}^{-1}\bq)
    \label{shift}
\eeq
where the configuration-space expression $\mathbf{\tilde S}_{*}\bP \equiv \mathbf{\tilde S} \ast \bP$ is short for a convolution corresponding to the multiplication in Fourier space.
We note that in terms of the argument $ \bk'=\mathbf{A}\bk$ of the Fourier transform of $\bP$, $\mathbf{S}$ corresponds to 
\beq
\mathbf{S}(\bk' ) =\left[ \frac{  k^{'2} S(\mathbf{A}^{-1} \bk')}{|\mathbf{A}^{-1} \bk' |^2}    \mathbf{A}^{-1}  \mathbf{A}^{-1}\right]
\equiv f(\bk')\mathbf{M},
\eeq
where we define $\mathbf{M} =\mathbf{A}^{-1} \mathbf{A}^{-1} $ and $f(\bk') =  k'^{2} S(\mathbf{A}^{-1} \bk') / |\mathbf{A}^{-1} \bk' |^2$.

The displaced and shifted fields now become
\beq
\delta^f_s(\bk) = \int d^3q e^{-i \bk \cdot \bq} \left( e^{-i \bk\cdot\left[ - \mathbf{A}\mathbf{\tilde S}_{*}  \bP(\mathbf{A}^{-1}  \bq)\right]} -1\right)
\eeq
and
\beq
\delta^f_d(\bk) = \int d^3q e^{-i \bk \cdot \bq} \left( e^{-i \bk \cdot \left[ \mathbf{A} \bP(\mathbf{A}^{-1}  \bq)  -\mathbf{A} \mathbf{\tilde S}_{*}\bP(\mathbf{A}^{-1}  \bq)\right]} -1\right). 
\label{displace}
\eeq
We back-substitute $\tilde \bq = \mathbf{A}^{-1} \bq$, relabel $\tilde{\mathbf{q}} \rightarrow \mathbf{q}$ and use $\bk' = \mathbf{A} \bk$ to obtain
\beq
\delta_s^{f}(\bk) = \mathrm{det}(\mathbf{A})\int d^3q\ e^{-i  \bk'\cdot\bq} \left( e^{-i \bk' \cdot\left[-\mathbf{\tilde S}_{*}\bP(\bq)\right]} -1\right) \label{squash1}
\eeq
and
\beq
 \delta_d^{f}(\bk) = \mathrm{det}(\mathbf{A})\int d^3q\ e^{-i\bk'\cdot\bq} \left( e^{-i\bk'\cdot\left[\bP(\bq)  - \mathbf{\tilde S}_{*}\bP(\bq)\right]} -1\right), \label{squash2}
 \quad
\eeq
noting that $\bk'= \mathbf{A}\bk$ is the conjugate wavenumber of the true, physical coordinate $\bx'=\mathbf{A^{-1}} \bx$.

These expressions were all calculated in the observed (unprimed, $\bx$ or $\bq$) coordinate system; however, as an aside, we note that the derivation can also be performed within the physical (primed) coordinate system, related by $\bq'=\mathbf{A^{-1}} \bq$. In this case we simply transform $\mathbf{s}^f(\bq)$ into the physical ($\bq'$) coordinate system via $\mathbf{s}^f_i(\bq) \rightarrow (\partial q'_i/\partial q_j) \mathbf{s}^f_j(\bq(\bq'))$; therefore, in the physical coordinate system, the ``wrong'' shift vector is $\mathbf{A^{-1}}\mathbf{s}^f(\mathbf{A q'})$. From Eq.~(\ref{shift}) we conclude that this equals $-\mathbf{\tilde S}_{*}\bP(\bq')$; Eqs.~(\ref{squash1} - \ref{squash2}) follow.

The reconstructed power spectrum, $P_{\mathrm{rec}}^f=P_{dd}+P_{ss}-2P_{ds}$, requires evaluation of 
\beqn
P_{ss}&\sim&\int d^3q\ e^{-i\bk'\cdot\bq}\left\langle e^{-i \bk' \cdot \left[ \mathbf{\tilde S}_{*}  \bP(\bq_1)-\mathbf{\tilde S}_{*} \bP(\bq_2) \right]} \right\rangle \nonumber \\&=& \int d^3q\ e^{-i\bk'\cdot\bq}\exp\left[-\frac{1}{2}\left\langle  \left( \bk' \cdot  \mathbf{\tilde S}_{*}\bP( \bq_1)-\bk' \cdot \mathbf{\tilde S}_{*}\bP(\bq_2) \right)^2 \right\rangle \right]
\eeqn
as well as similar terms for $P_{dd}$ and $P_{ds}$. Expanding out the square in the exponential gives a zero-lag part (involving products of $\bP$ evaluated at the same position) and a non-zero-lag piece (involving products evaluated at different positions $\bq_1, \bq_2$). We will evaluate these in turn. The zero-lag part of the $P_{ss}$ term is
\begin{align}
&\hphantom{=} \exp \left[-\left\langle \sum_{ij} k'_i  [\mathbf{\tilde{S}}_{*}\bP(0) ]_i  k'_j [\mathbf{\tilde{S}}_{*}\bP(0) ]_j \right\rangle \right] \nonumber \\
&= \exp \left[-\sum_{ij} k'_i k'_j \int \int \frac{d \bk_1 d \bk_2}{(2\pi)^6} \left[ -i  \frac{\mathbf{M} \bk_1}{k_1^2} \right]_i   \left[ -i  \frac{\mathbf{M} \bk_2}{k_2^2} \right]_j f(\bk_1) f(\bk_2) \langle \delta(\bk_1) \delta(\bk_2) \rangle \right]\nonumber \\ 
&=\exp \left[- \sum_{i} k_i^{'2} \int   \frac{d \bk_1}{(2\pi)^3} \left[ \mathbf{M} \bk_1 \right]_i^2  f^2(\bk_1)  \frac{P_L(k_1)}{k_1^4} \right]\nonumber \\
&\equiv \exp\left[ - \frac{1}{2}\sum_i k_i^{'2} \Sigma^2_{ss,i} \right]
\end{align}
where we note that $\left[ \mathbf{M} \bk_1 \right]_i=k_{1,i}/\alpha_i^2$ and we have defined $\Sigma^2_{ss,i}$ from the previous line.

For the analogous zero-lag terms of $P_{dd}$ and $P_{ds}$, we obtain
\beqn 
\exp \left[- \sum_{i} k_i^{'2} \int\frac{d \bk_1}{(2\pi)^3} \left[ \{1-f(\bk_1)\mathbf{M}\} \bk_1 \right]_i^2   \frac{P_L(k_1)}{k_1^4} \right]
\equiv \exp\left[ - \frac{1}{2}\sum_i k_i^{'2} \Sigma^2_{dd,i}(\mathbf{A})  \right]
\eeqn
and
\beqn 
 \exp\left[ -\frac{1}{2}\sum_i k_i^{'2} \frac{\Sigma^2_{dd,i} (\mathbf{A}) + \Sigma^2_{ss,i} (\mathbf{A})}{2} \right]
\equiv \exp\left[ - \frac{1}{2}\sum_i k_i^{'2} \Sigma^2_{ds,i} (\mathbf{A}) \right]
\eeqn
by similar manipulations.

We now discuss the non-zero-lag pieces. After expanding the exponential, the non-zero-lag part of the $P_{ss}$ term is 
\begin{align}
& \int d^3q\ e^{-i\bk'\cdot\bq} \sum_{ij}k'_i k'_j \left\langle[\mathbf{\tilde{S}}_{*}\bP(\bq_1)]_i [\mathbf{\tilde{S}}_{*}\bP(\bq_2)]_j \right\rangle \\
&\approx \int d^3q\ e^{-i\bk'\cdot\bq} \int\frac{d^3p}{(2\pi)^3}\ e^{i\mathbf{p}\cdot\bq} \sum_{ij}k'_i k'_j
\frac{[\mathbf{M p}]_i[\mathbf{Mp}]_jf^2(\mathbf{p})}{p^4}
P_L(p) \\
&= \frac{[\bk'\cdot\mathbf{Mk'}]^2f^2(\bk')}{k^{'4}} P_L(k')
\end{align}
Similarly, for the $P_{dd}$ and $P_{ds}$ terms we obtain
\beqn
\frac{\left[\bk'\cdot\{\mathbf{1}-f(\bk')\mathbf{M} \}\bk' \right]^2}{k^{'4}} P_L(k')
\quad {\rm and}\quad
 -\frac{\left[\bk'\cdot \{\mathbf{1}-f(\bk')\mathbf{M} \}  \bk' \right]\left[\bk'\cdot \mathbf{M}\bk' f(\bk')\right]}{k^{'4}} P_L(k') \quad .
\eeqn

Summarizing and noting that $\mathbf{k'\cdot Mk'}=|\mathbf{A^{-1} k'}|^2$ so that $\mathbf{k' \cdot Mk'}f(\bk') = k'^2S(\mathbf{A}^{-1}\bk')=k'^2 S(k)$, while $\delta^{(D)}(\mathbf{A k})=\delta^{(D)}(\mathbf{k})/\mathrm{det}(\mathbf{A})$, we obtain: 
\beqn
P_{\mathrm{rec}}^f(\bk) &=& \mathrm{det}(\mathbf{A}) P_L(\bk')
\left\{\vphantom{\int} \right. \nonumber \\
&&\hphantom{+} \exp\left[ -\frac{1}{2}\sum_i k_i^{'2} \Sigma^2_{dd,i} (\mathbf{A})\right]  \left[1-S(k) \right]^2 \nonumber \\
&&+ \exp\left[ -\frac{1}{2}\sum_i k_i^{'2} \Sigma^2_{ss,i}(\mathbf{A})\right]  \left[S(k) \right]^2 \nonumber \\
&&+\left. \exp\left[ -\frac{1}{2}\sum_i k_i^{'2} \Sigma^2_{ds,i}(\mathbf{A})\right]
2 S(k) \left[1-S(k)\right]
 \right\}
\eeqn
This expression is our final result for the case of reconstruction with incorrect distances. We note that the results are similar in form to Eqs. (\ref{simple1}-\ref{simple2}); we recover these standard results for $\mathbf{A} =\mathbf{1}$, as expected.

We note that for $\mathbf{A} =\mathbf{1}$, we recover the standard results of Eqs. (\ref{simple1}-\ref{simple2}) as expected.

\section{Incorrect distances, bias and growth rate}
\label{sec:everything_wrong}
We now turn to the full problem: reconstruction with the wrong bias and redshift space distortion parameters, in addition to incorrect fiducial distances. Our starting point here is the density field, again measured with wrong assumptions about the distances, but now including bias and redshift space distortions as in Section \ref{sec:background}. Following an argument as in the previous section, we obtain (for $\mathbf{k}\ne 0$):
\beq
 \delta^{f}(\bk) = \mathrm{det}(\mathbf{A}) \int\frac{d\lambda}{2\pi}\,d^3q\  F(\lambda) e^{-i\bk'\cdot\bq+i\lambda\delta(\bq)} e^{-i \bk'\cdot \left[\mathbf{R}\bP(\bq)\right]}\label{eeq2},
\eeq
with $R_{ij}=\left(\delta_{ij}+{f}\hat{z}_i\hat{z}_j\right)$. 

We now discuss reconstruction. In our analysis we consider a case where the growth rate is in fact $f$, but we incorrectly assume it to be $\tilde f$.  We define $\tilde{R}_{ij}=\left(\delta_{ij}+\tilde{f}\hat{z}_i\hat{z}_j\right)$.
Similarly the true bias is $b$ but we incorrectly assume it to be $\tilde{b}$.  The negative Zel'dovich displacement we infer is then
\beq
\mathbf{s}^f(\bk) = - i\alpha_\perp^2 \alpha_\parallel \frac{\bk}{k^2} S(k) \frac{\delta^{\rm obs}(\mathbf{A} \bk)}{{\tilde b+\tilde f\mu^2}} = - i\alpha_\perp^2 \alpha_\parallel \frac{\bk}{k^2} S(k)\ \frac{ b+ f\mu_{\mathbf{A}\bk}^2}{{\tilde b+\tilde f\mu^2}}\ \delta(\mathbf{A} \bk),
\label{densitydiv}
\eeq
where $\mu_{\mathbf{A\bk}}^2 = \left[(\mathbf{A} \bk) \cdot\mathbf{\hat z}/|\mathbf{A} \bk|\right]^2$.
Rewriting this in terms of the rescaled vector $\bk' = \mathbf{A} \bk$ as before and repeating the arguments following Eq.~(\ref{eqn:vecS_defn}), we obtain $\mathbf{s}^f(\bq)$ as defined previously (Eq.~\ref{shift}), except that now $\mathbf{S}$ is instead given by
\beq
\mathbf{S}(\bk' ) \equiv  \mathbf{A}^{-2}  \frac{ k^{'2}}{|\mathbf{A}^{-1} \bk' |^2} \left[   \frac{ b+ f\mu'^2}{{\tilde b+\tilde f\nu^2}}  \right]  S(\mathbf{A}^{-1} \bk') 
\eeq
where $\mu'=\mu_{\mathbf{A k}}=\bk' \cdot \hat{\mathbf{z}}/{k'}$ and we have defined $\nu^2\equiv\mu_{\mathbf{A^{-1}}\bk'}^2 = \left[(\mathbf{A^{-1}}\bk') \cdot\mathbf{\hat z}/|A^{-1}\bk'|\right]^2=\mu^2=(\hat{\mathbf{k}}\cdot\hat{\mathbf{z}})^2$ to emphasize the distinction between the two variables.

Displacing galaxies and randoms by $\mathbf{\tilde{R}}\mathbf{s}^f$, we obtain the shifted and displaced fields (assuming $\mathbf{k}\ne 0$):
\beq
\delta_s^{f}(\bk) = \mathrm{det}(\mathbf{A}) \int d^3q\ e^{-i\bk'\cdot\bq} e^{-i\bk'\cdot\left[-\mathbf{\tilde R}\mathbf{\tilde S}_{*}\bP(\bq)\right]}\label{eeq1}
\eeq
and
\beq
 \delta_d^{f}(\bk) = \mathrm{det}(\mathbf{A}) \int\frac{d\lambda}{2\pi}\,d^3q\  F(\lambda) e^{-i\bk'\cdot\bq+i\lambda\delta(\bq)} e^{-i \bk'\cdot \left[\mathbf{R}\bP(\bq) - \mathbf{\tilde R}\mathbf{\tilde S}_{*}\bP(\bq)\right]}\label{eeq2}
\eeq

Though Eqs.~(\ref{eeq1}-\ref{eeq2}) essentially follow as in the previous section, one small subtlety is that the biasing exponent $e^{i\lambda \delta(\mathbf{A^{-1}q})}$ could potentially complicate the displacement operation of Eq.~(\ref{displace}). However, our alternative derivation in the physical coordinate system proceeds exactly as previously, confirming the above expressions.

We may now proceed to calculating the incorrectly reconstructed power spectrum, following the procedure in the previous section closely. To simplify the exposition, we explain all the details of the calculation in Appendix A, and here simply quote the final results: we obtain a general equation for the reconstructed power spectrum with errors in angle, distance, bias, and RSD. 
The final expression is:
\begin{align}
P_{\mathrm{rec}}^f(\bk) &= \alpha_\bot^2 \alpha_\parallel \left[b+f\mu^{'2}\right]^2 P_L(k') \left\{
\left[ \frac{1+\tilde f\nu^2}{\tilde b+\tilde f\nu^2} S(k)\right]^2 \mathcal{D}_{ss} \right. \nonumber \\
&+ \left. \left[1-  \frac{1+ \tilde f \nu^2 }{\tilde b+ \tilde f \nu^2 } S(k) \right]^2 \mathcal{D}_{dd}
+  2\left[1 - \frac{1+\tilde f\nu^2}{\tilde b+\tilde f\nu^2} S(k)\right] \left[ \frac{1+\tilde f\nu^2}{\tilde b+\tilde f\nu^2} S(k)\right]  \mathcal{D}_{ds} \right\},
\label{eqn:Pfalse_rec}
\end{align}
where 
\begin{align}
-2\ln\mathcal{D}_{ss} &= k_\bot^{'2} \Sigma^{{2~\tilde f, \tilde b, \mathbf{A}}}_{ss,\bot} + k_\parallel^{'2} (1+\tilde f)^2 \Sigma^{{2~\tilde f, \tilde b, \mathbf{A}}}_{ss,\parallel} \\
-2\ln\mathcal{D}_{dd} &= k_\bot^{'2} \Sigma^{{2~\tilde f, \tilde b, \mathbf{A}}}_{dd,\bot} + k_\parallel^{'2} (1+f)^2 \Sigma^{{2~\tilde f, \tilde b, \mathbf{A}}}_{dd,\parallel} \\
-2\ln\mathcal{D}_{ds} &= \frac{1}{2}\left[
-2\ln\mathcal{D}_{dd} - 2\ln\mathcal{D}_{ss} \right],
\end{align}
where we remind the reader that $\bk =\mathbf{A^{-1}k'}$ and $\nu=\hat{\bk}\cdot \hat{\mathbf{z}}$. For convenience, we will later refer to the expression in curly brackets as the damping function $D(k,\mu, \mathbf{A},\tilde{f},\tilde{b},f,b)$. The relevant damping coefficients are given by:
\beqn 
\frac{1}{2} \Sigma^{2~\tilde f, \tilde b, \mathbf{A}}_{ss,\bot} = \int\frac{d \bk_1}{(2\pi)^3} (1-\mu_1^2)k_{1}^2  g^2(\bk_1)S^2(\mathbf{A^{-1}} \bk_1) \frac{P_L(k_1)}{2 k_1^4 \alpha_\bot^4 }\nonumber,
\label{defS1}
\eeqn
\beq
\frac{1}{2} \Sigma^{2~\tilde f, \tilde b, \mathbf{A}}_{ss,\parallel} =  \int\frac{d\bk_1}{(2\pi)^3} \mu_1^2 k_{1}^2 g^2(\bk_1)S^2(\mathbf{A^{-1}} \bk_1) \frac{P_L(k_1)}{k_1^4 \alpha_\parallel^4 }\nonumber
\eeq
and
\beqn 
\frac{1}{2}\Sigma^{2~\tilde f, \tilde b, \mathbf{A}}_{dd,\bot} =  \int   \frac{d \bk_1}{(2\pi)^3}  (1-\mu^2) k_{1}^2 (1- \frac{g(\bk_1)S(\mathbf{A^{-1}}\bk_1)}{\alpha_\bot^2})^2\frac{P_L(k_1)}{2 k_1^4 }
\eeqn
\beqn
\frac{1}{2} \Sigma^{2~\tilde f, \tilde b, \mathbf{A}}_{dd,\parallel} = \int\frac{d \bk_1}{(2\pi)^3} \mu^2 k_{1}^2 (1- \frac{1+\tilde f}{(1+ f)\alpha_\parallel^2}g(\bk_1)S(\mathbf{A^{-1}}\bk_1))^2 \frac{P_L(k_1)}{k_1^4  },
\eeqn
where $g(\bk') = \left[\frac{ k^{'2}}{|\mathbf{A}^{-1} \bk' |^2}   \frac{ b+ f\mu'^2}{{\tilde b+\tilde f\nu^2}}  \right]$.

We note that, aside from errors in angle, distance, bias, and RSD, other wrong assumptions in cosmological parameters could occur; however, as long as these do not affect the construction of the displacement in Eq.~(\ref{densitydiv}), they do not have any reconstruction-specific effects. For example, even if the wrong linear power spectrum shape was assumed, the measured power spectrum would still be well-approximated by Eq.~(\ref{eqn:Pfalse_rec}) with the true $P_L(k)$.
\section{Discussion}
\label{sec:discussion}

To gain an understanding of the impact of incorrect BAO reconstruction, we will now evaluate the expressions derived above for different types of wrong assumptions made in reconstruction. In particular, we will evaluate the multipoles of the wrongly-reconstructed power spectrum
\beq
P_\ell(k) = \frac{2\ell+1}{2}\int_{-1}^{1} d\mu\ P^f_{\mathrm{rec}}(k,\mu)\mathcal{L}_\ell(\mu)
\eeq
where $\mathcal{L}_\ell$ is the Legendre polynomial of order $\ell$. We will consider the following errors, each in turn: (i) a wrong scaling perpendicular to the line of sight, by $\alpha_\perp = 1.03$; (ii) an incorrect scaling along the line of sight, by $\alpha_\parallel = 1.03$ (iii) an incorrect bias of $b=1.1\times 2$ (the actual bias is taken to be $b=2$) (iv) an incorrect RSD parameter of $f=1.3\times 0.55$ (where the correct $f=0.55$). A $3\%$ error  in distances was chosen because it corresponds approximately to the difference between the Planck and WMAP cosmologies evaluated at the redshifts of the BOSS survey. To calculate the matter power spectrum, we will use the fitting formula of \cite{EisHu}, assuming the TT, TE, EE + lowP + lensing + ext cosmology of the 2015 Planck release \cite{PlanckParams}. We will further assume a Gaussian smoothing function $S(k)=e^{-k^2\Sigma_{\mathrm{sm}}^2/4}$ following \cite{PWC09}, with a default smoothing scale of $\Sigma_{\mathrm{sm}}=14.1$ Mpc/h, though we will also discuss the impact of other choices.

\begin{figure*}
     \centering \subfigure[Wrong $\alpha_\perp$=1.03
     ]{\includegraphics[width=.48\textwidth,clip]{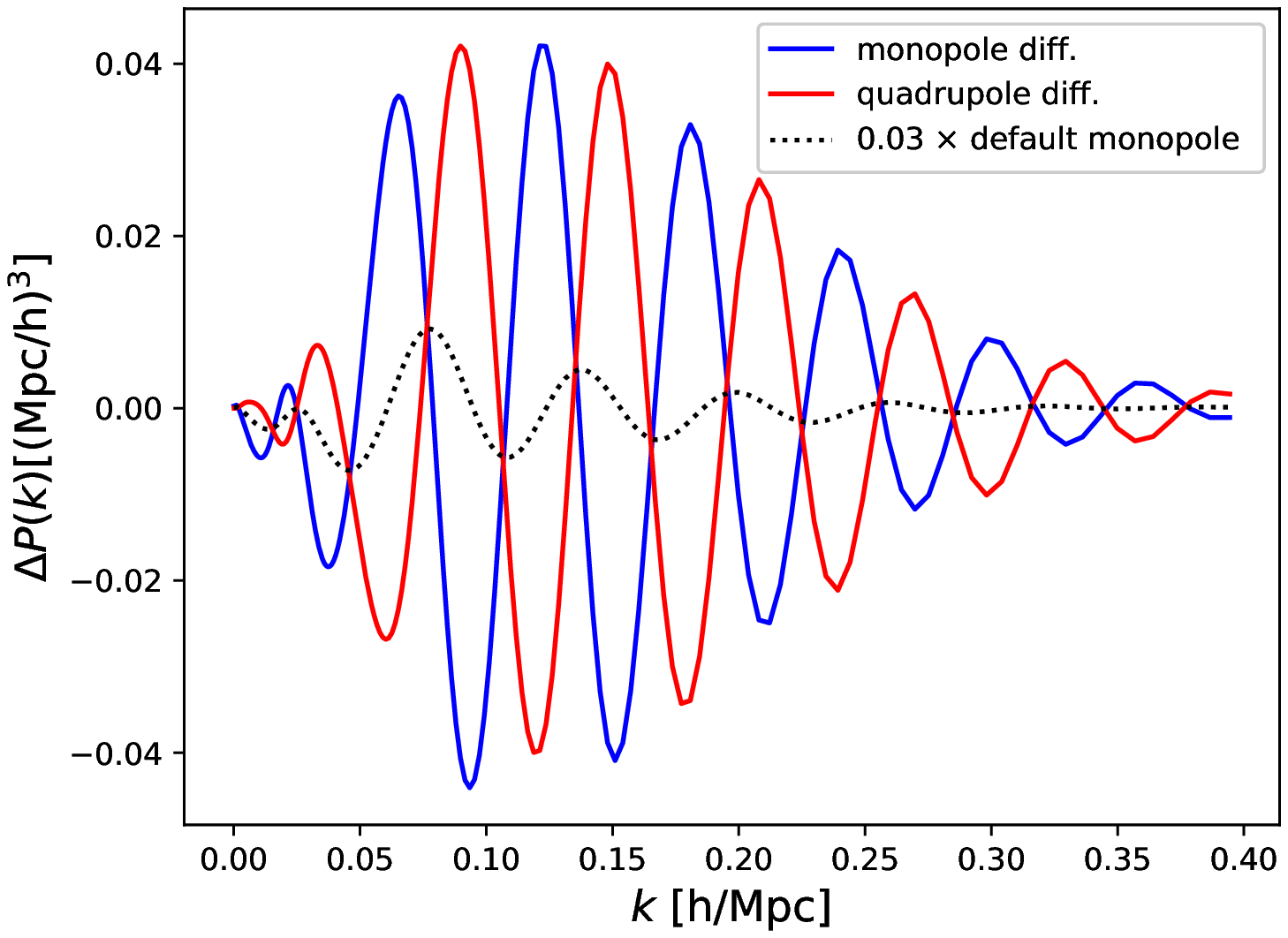}}
     \subfigure[Wrong $\alpha_\parallel$=1.03
     ]{\includegraphics[width=.48\textwidth,clip]{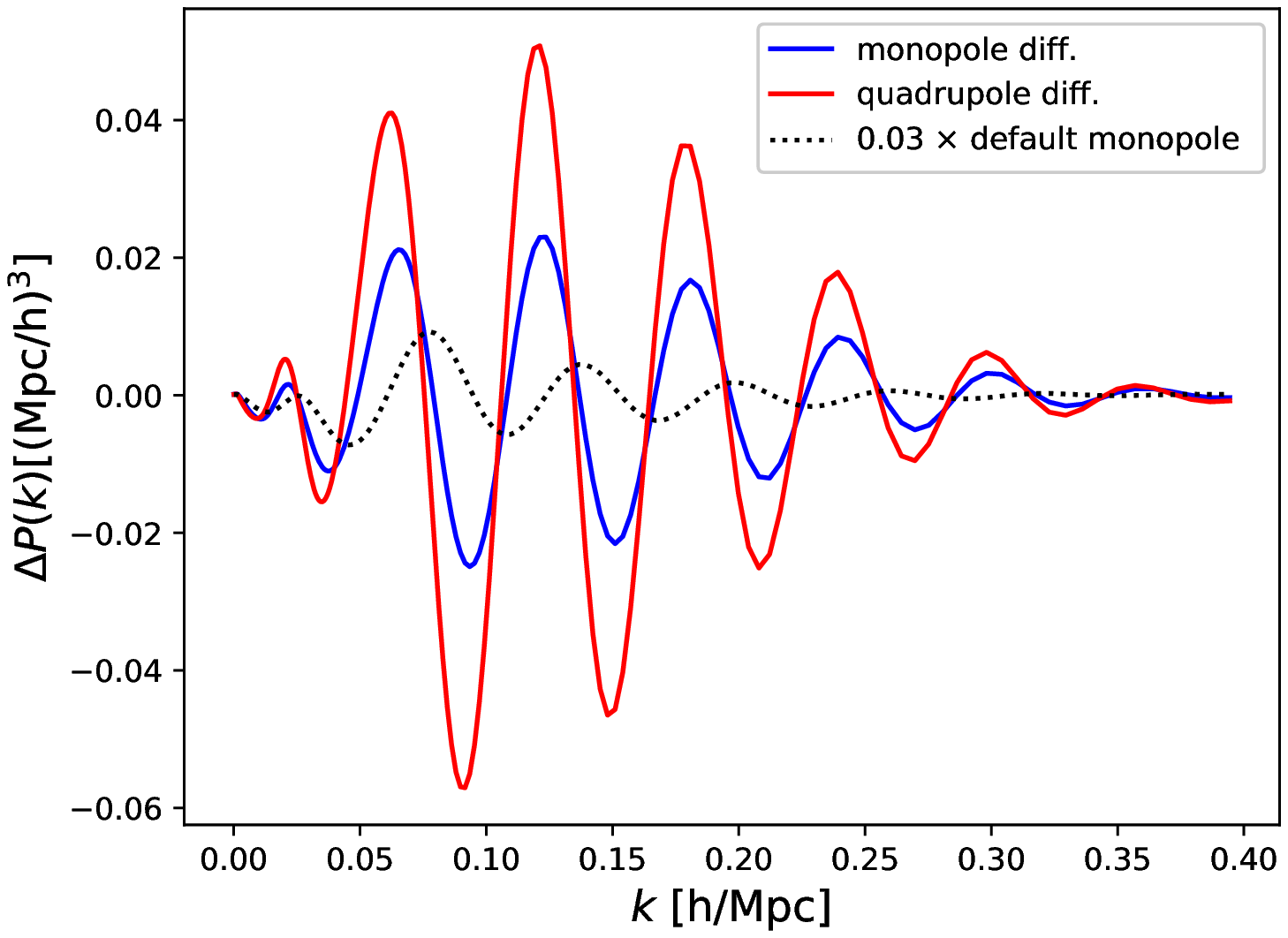}}\\
     \subfigure[Wrong $b=1.1\times 2$
     ]{\includegraphics[width=.48\textwidth,clip]{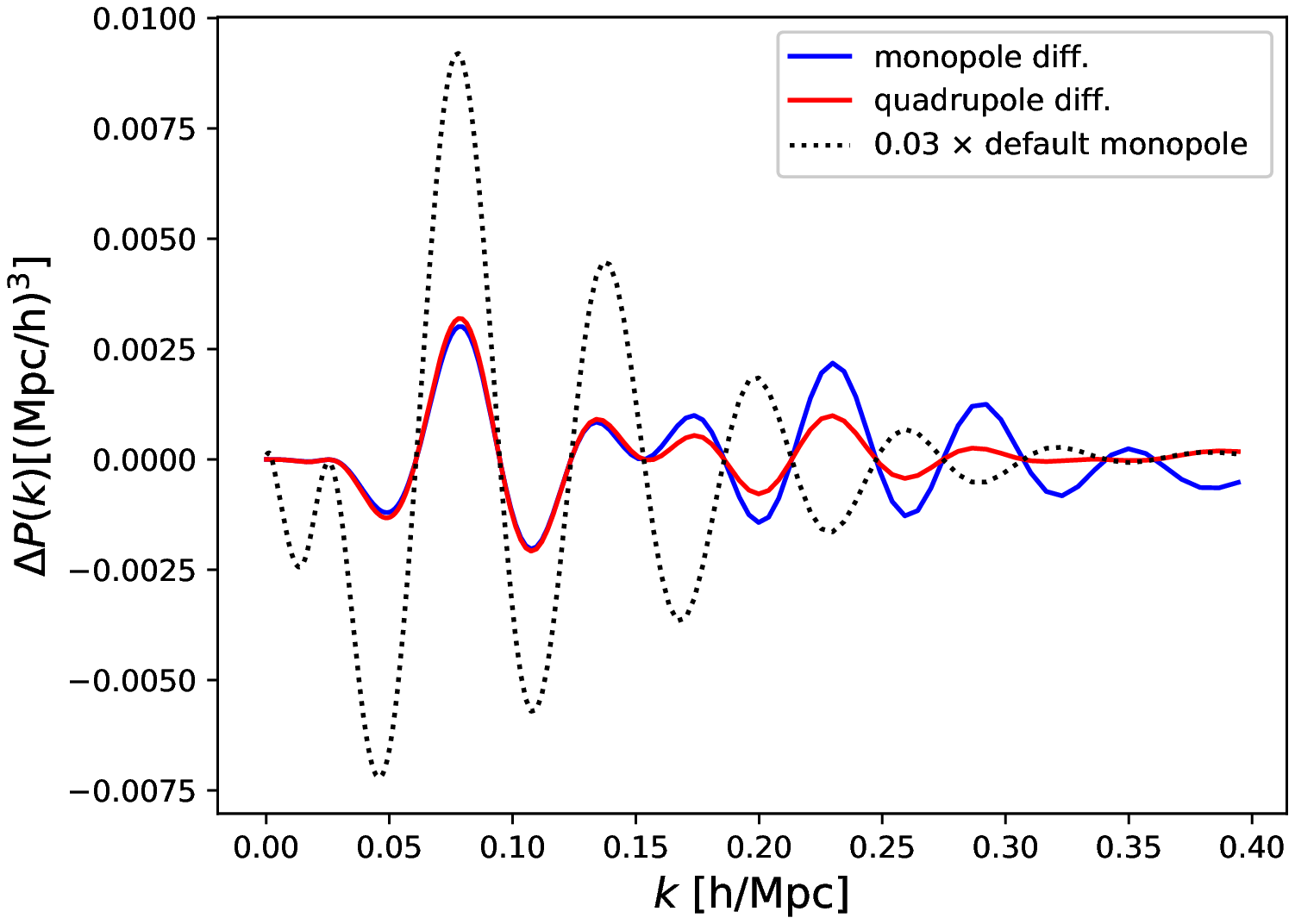}}
     \subfigure[Wrong $f=1.3\times 0.55$
     ]{\includegraphics[width=.48\textwidth,clip]{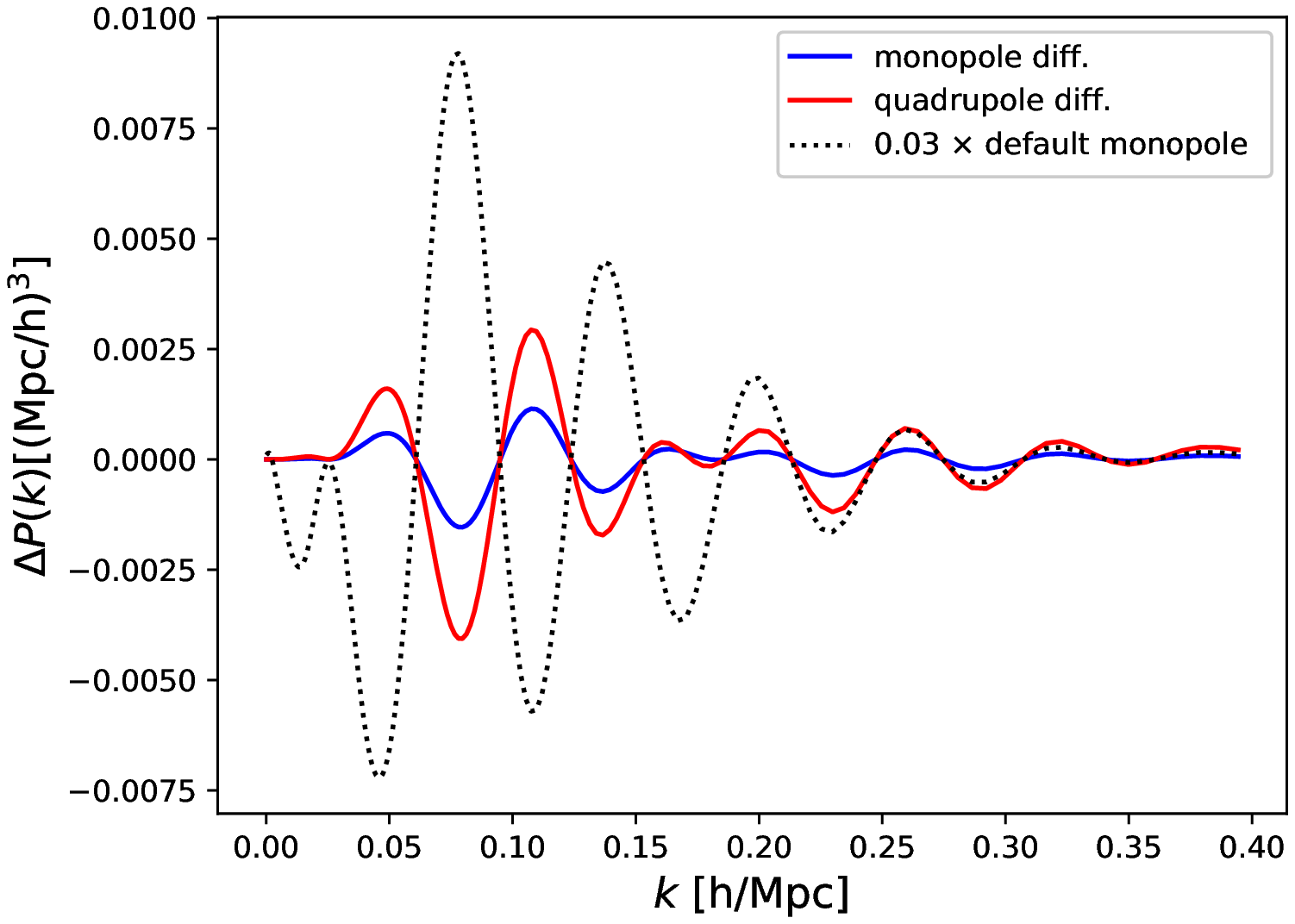}}\\
     \label{tbf} \caption{Effect of errors in fiducial cosmology, bias and RSD on the reconstructed power spectra. Shown are the differences between the incorrectly reconstructed power spectra and the correctly reconstructed default ones; the monopole difference is plotted in blue, the quadrupole difference in red. For reference, the default reconstructed power spectrum monopole itself is shown, rescaled by a factor $0.03$, with a black dotted line. (Note that for all power spectra we have isolated the BAO feature by first subtracting a ``no-wiggle'' power spectrum without oscillations \cite{EisHu} and then dividing by the linear power spectrum amplitude.)}
\label{fig:Pk_all_diff}
\end{figure*}

We begin by considering the Fourier space spectra obtained by calculating the multipoles of Eq.~(\ref{eqn:Pfalse_rec}). For different types of `wrong reconstruction', we show the differences between the incorrectly reconstructed monopole and quadrupole power spectra and the correctly reconstructed versions in Fig.~\ref{fig:Pk_all_diff}. 

For errors in $f$ and $b$, changes of order 1\% in the monopole are seen on some scales. Comparing with the default monopole, it can be seen that the incorrect reconstruction spectrum is approximately related to the default reconstruction spectrum by a transfer function that changes the oscillation envelope of the BAO peaks for both monopole and quadrupole, but does not change the peak locations significantly.

A similar conclusion -- that the peak positions are not significantly affected by incorrect reconstruction -- applies when considering errors in cosmology, i.e. $\alpha_\perp \neq 1$ or $\alpha_\parallel \neq 1$; however, it should be noted that in this case, we should not compare the BAO peak scale with the $\alpha=1$ cases, but instead must search for the ``extra'' shift from incorrect reconstruction by comparing the measured scale with the rescaled correctly reconstructed power spectrum $P(\bk')$\footnote{The rescaled correctly reconstructed power spectrum is obtained by simply substituting $k, \nu$ by $k', \mu'$, and preserving the normalizing prefactor $\alpha_\parallel \alpha_\perp^2$ for simpler comparisons.}.

We now approximately quantify the BAO monopole scale shift in Fourier space; rather than perform a fit for the BAO scale for a certain experimental configuration, we simply use the average of the first peak and trough locations as a rough estimate for the BAO scale (noting that higher peaks give consistent results). For the given errors in $\alpha_\perp$, $\alpha_\parallel$, $b$, $f$ we find monopole peak position shifts of order $\sim 3 \times 10^{-4}$ or less compared to the rescaled correctly reconstructed spectrum\footnote{An explanation for the non-zero (but still negligible) shifts found for $\alpha \neq 1$ could be that the smooth functions multiplying the oscillations (such as the damping factors) are modified by incorrect reconstruction in a way that is anisotropic; therefore, the relative contributions of perpendicular and parallel BAO scales to the BAO monopoles and quadrupoles is changed, resulting in a non-zero shift in the peak positions. This -- already negligible -- effect could be further reduced by fitting different damping scales in the line-of-sight-perpendicular and parallel directions; or perhaps to some extent by fitting a mean damping scale, as is typically done.}. This difference is negligible for experiments operating in the foreseeable future, and likely smaller than shifts induced by terms we have neglected. 

\begin{figure*}
     \centering \subfigure[Wrong $\alpha_\perp$=1.01
     ]{\includegraphics[width=.48\textwidth,clip]{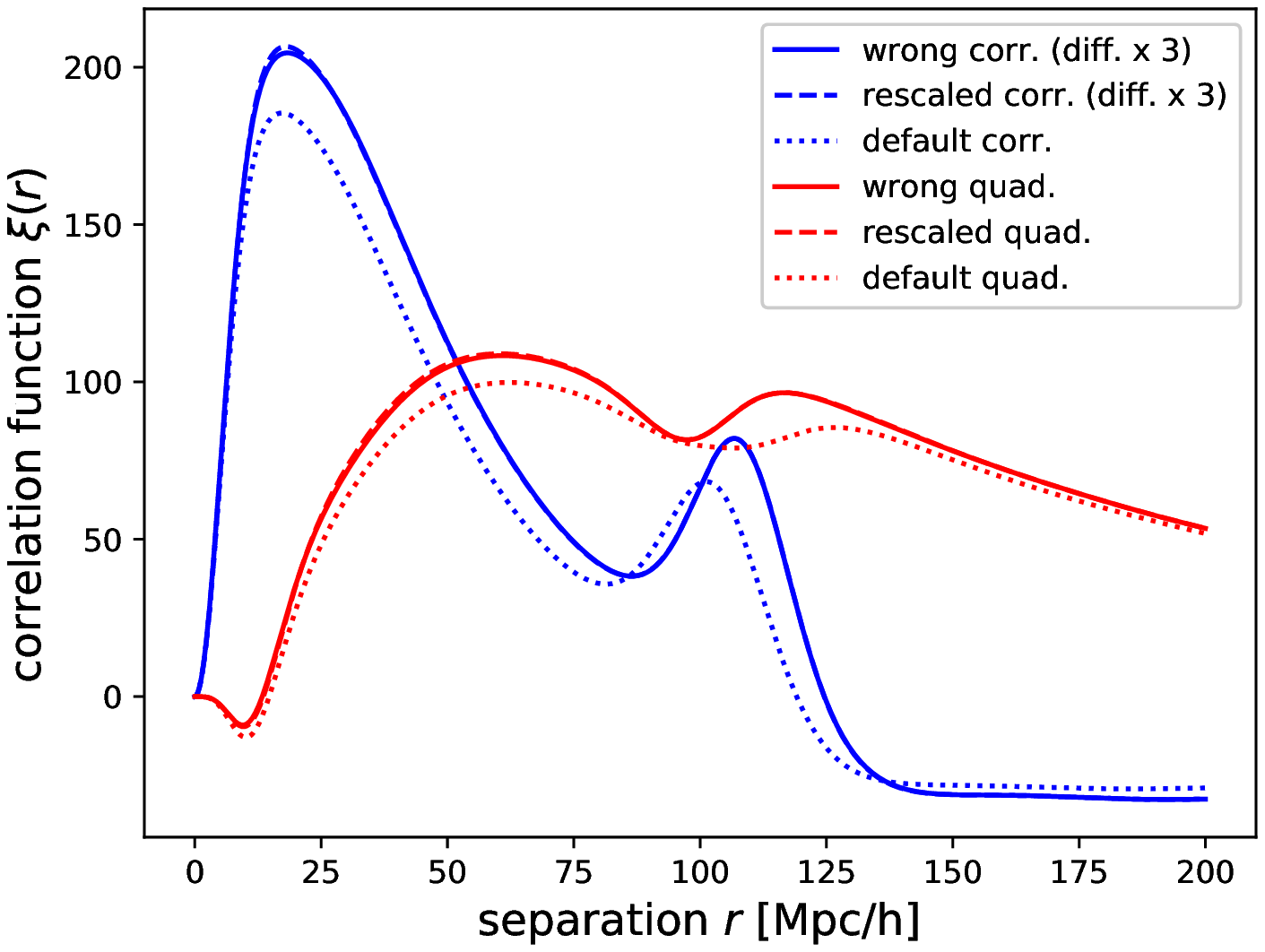}}
     \subfigure[Wrong $\alpha_\parallel$=1.01
     ]{\includegraphics[width=.48\textwidth,clip]{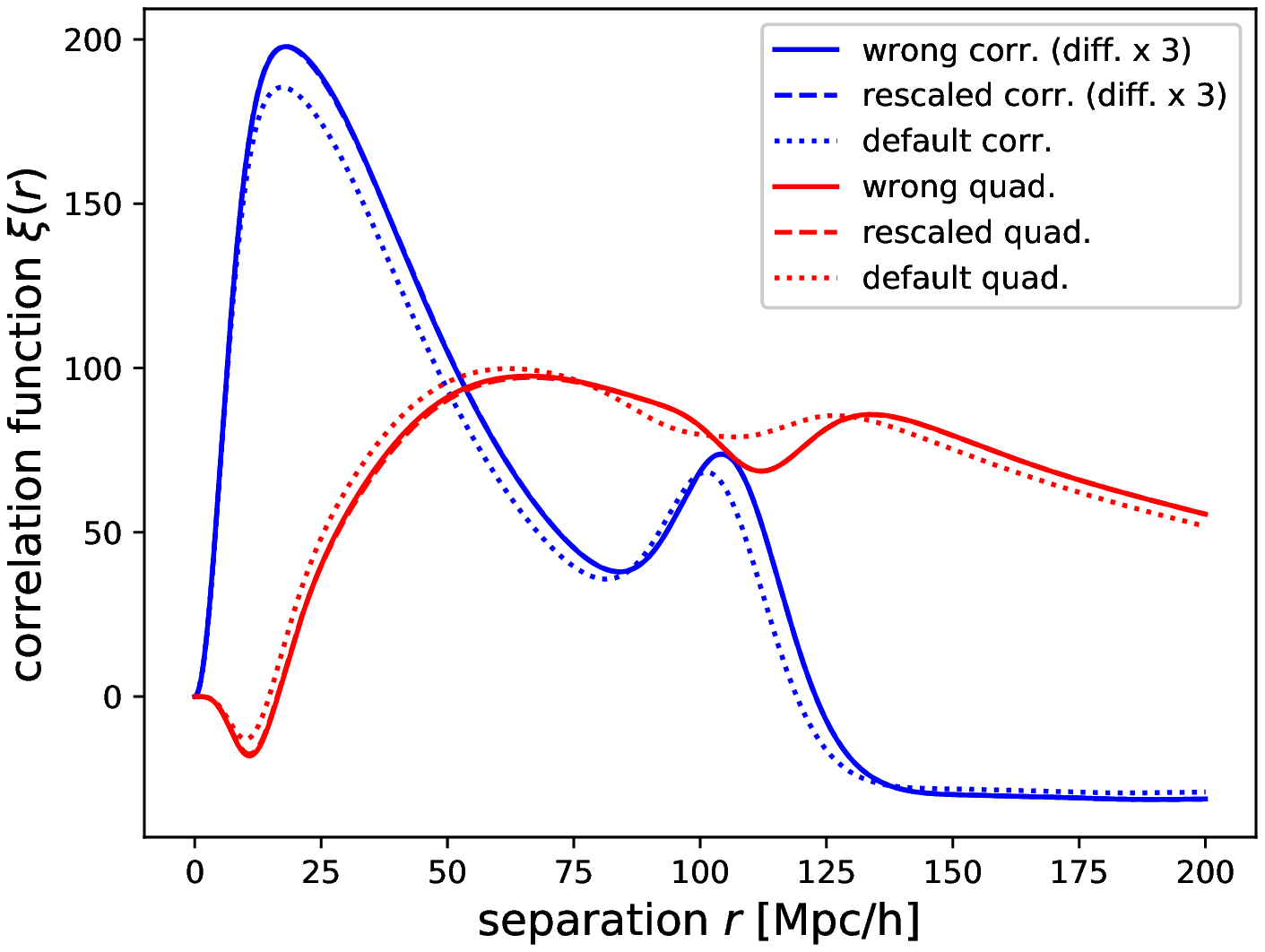}}\\
     \subfigure[Wrong $b=1.1\times 2$
     ]{\includegraphics[width=.48\textwidth,clip]{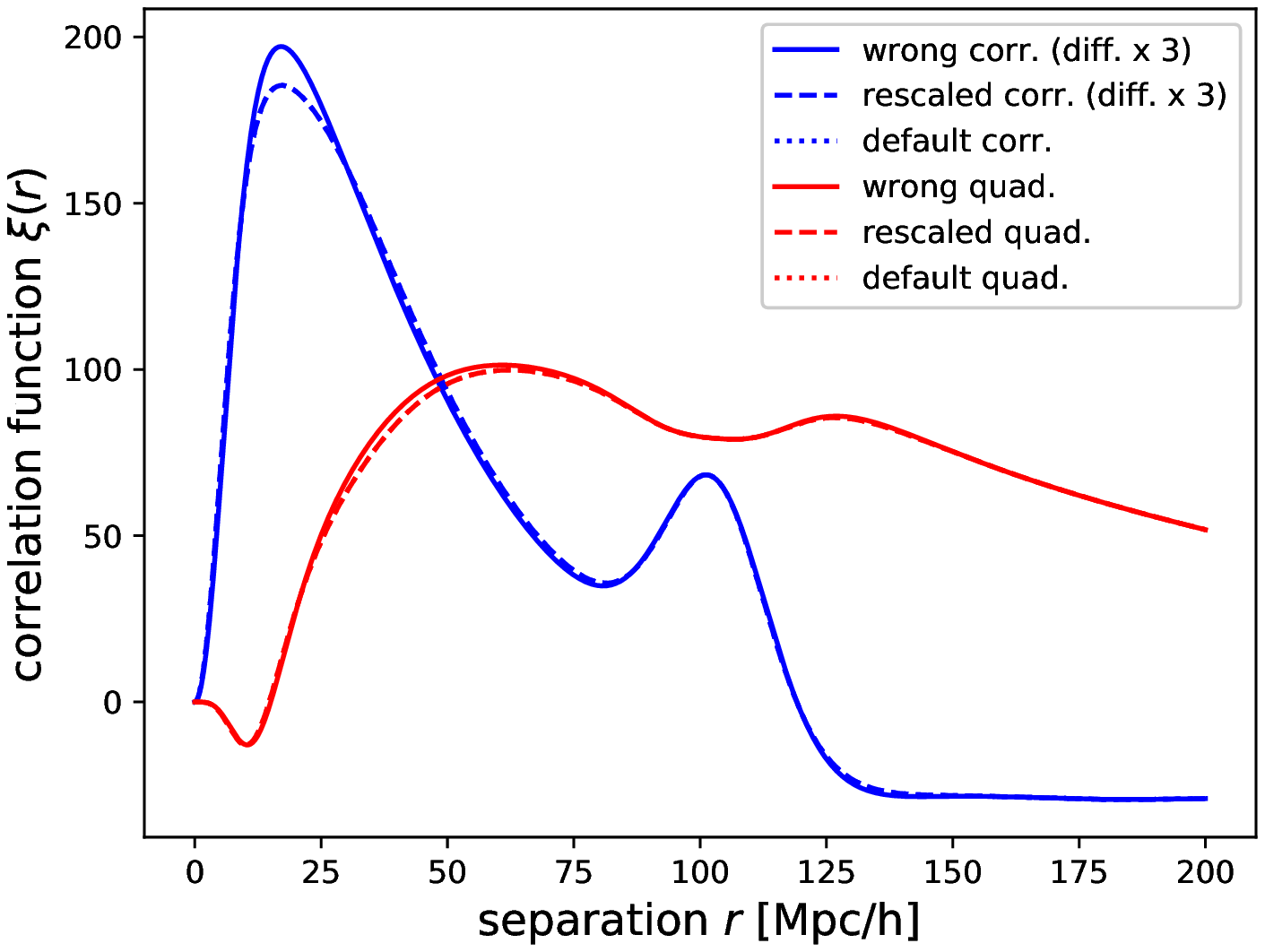}}
     \subfigure[Wrong $f=1.3\times 0.55$
     ]{\includegraphics[width=.48\textwidth,clip]{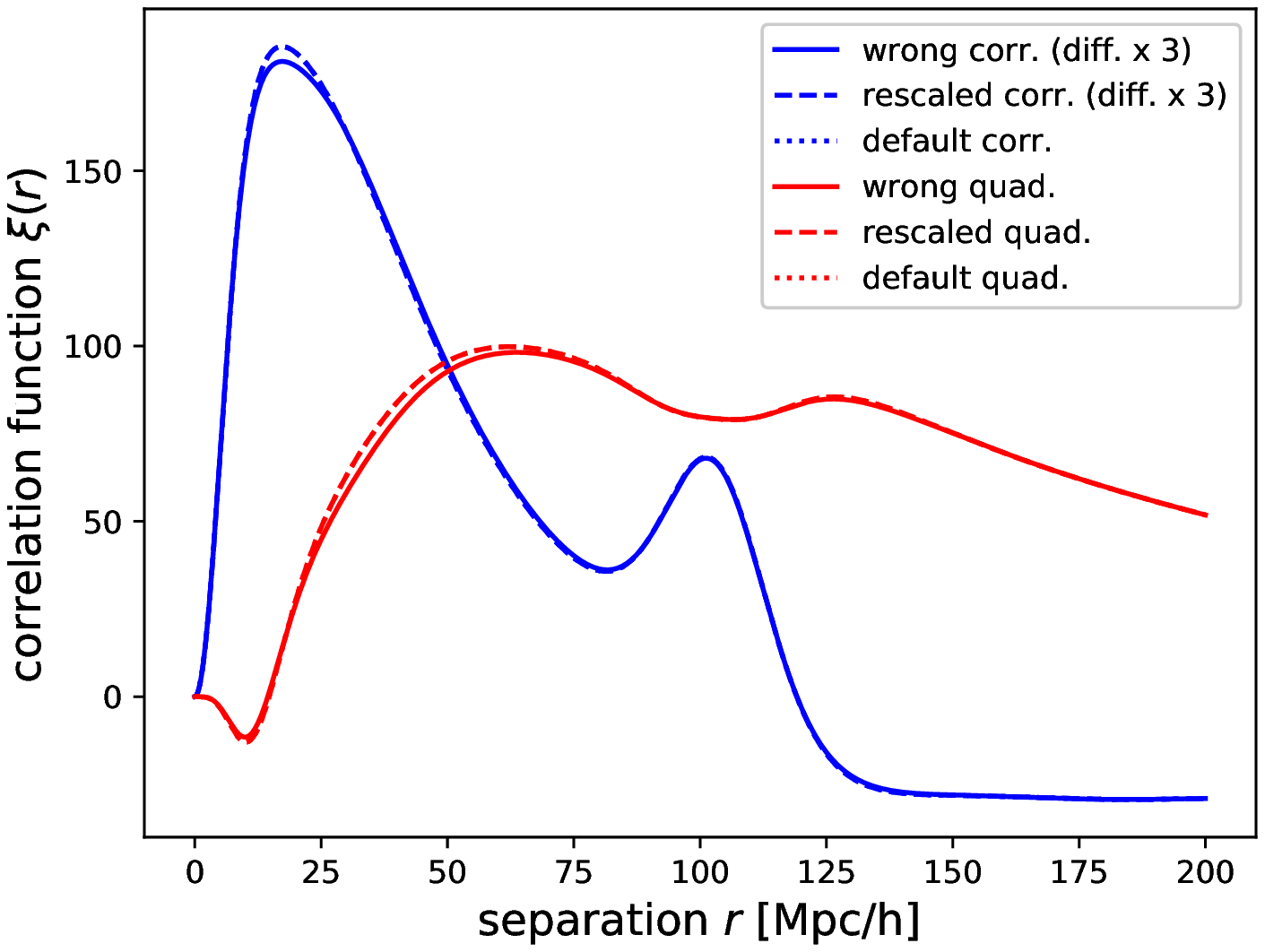}}\\
     \label{tbf} \caption{Configuration space picture of wrong reconstruction. Shown are the wrongly reconstructed correlation function (solid line), the rescaled correctly reconstructed correlation function (with $\bk\rightarrow \mathbf{A}\mathbf{k}$; dashed line) and the default correctly reconstructed correlation function (dotted line). To show the differences from the default reconstruction more clearly, we add twice the difference with respect to the default reconstruction to the wrong and rescaled curves. The monopole is shown in blue, the quadrupole in red.}
\label{fig:Xi_all}
\end{figure*}

\begin{figure*}
     \centering \subfigure[Wrong $\alpha_\perp$=1.03: ratio
     ]{\includegraphics[width=.48\textwidth,clip]{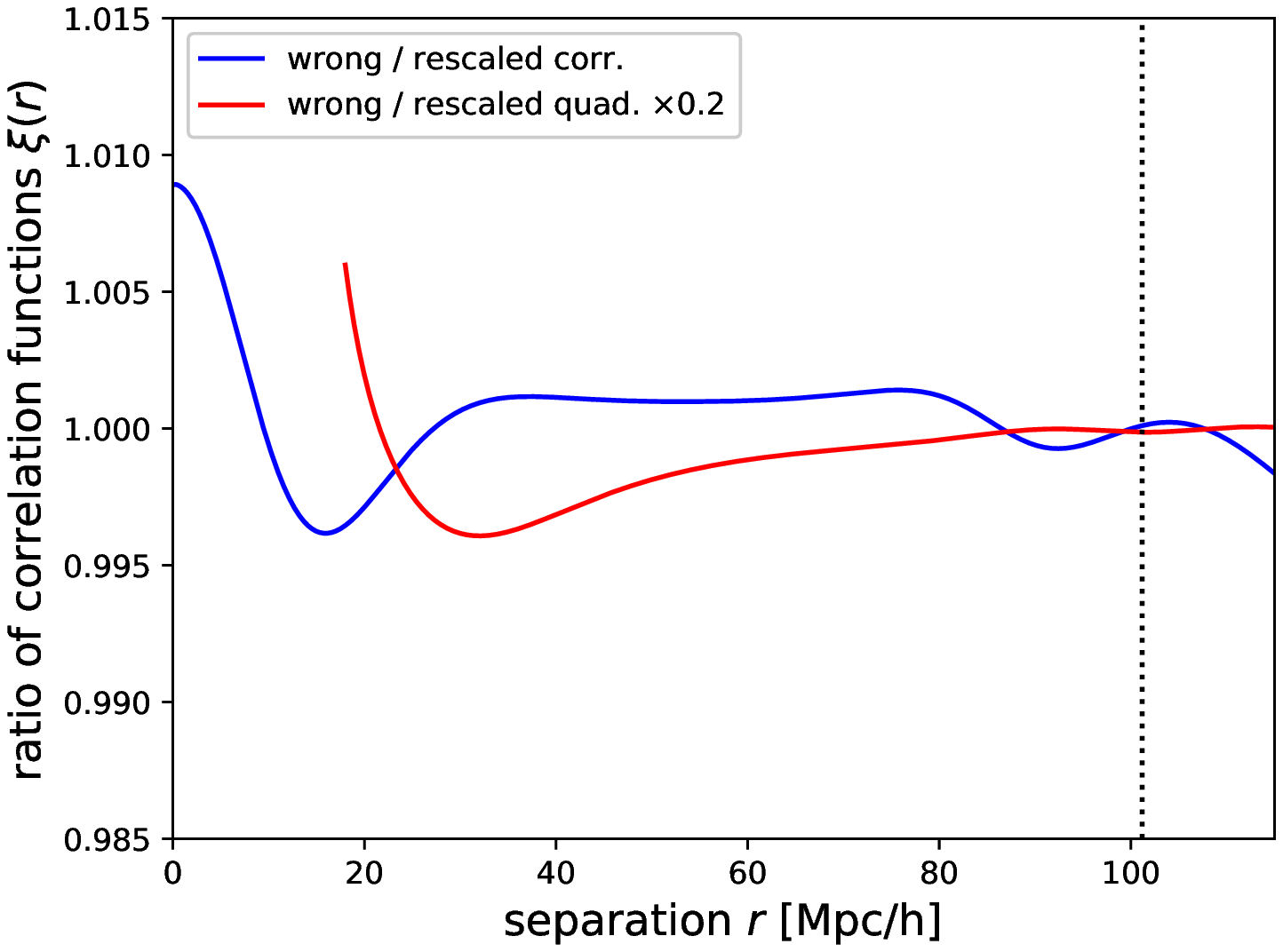}}
     \subfigure[Wrong $\alpha_\parallel$=1.03: ratio
     ]{\includegraphics[width=.48\textwidth,clip]{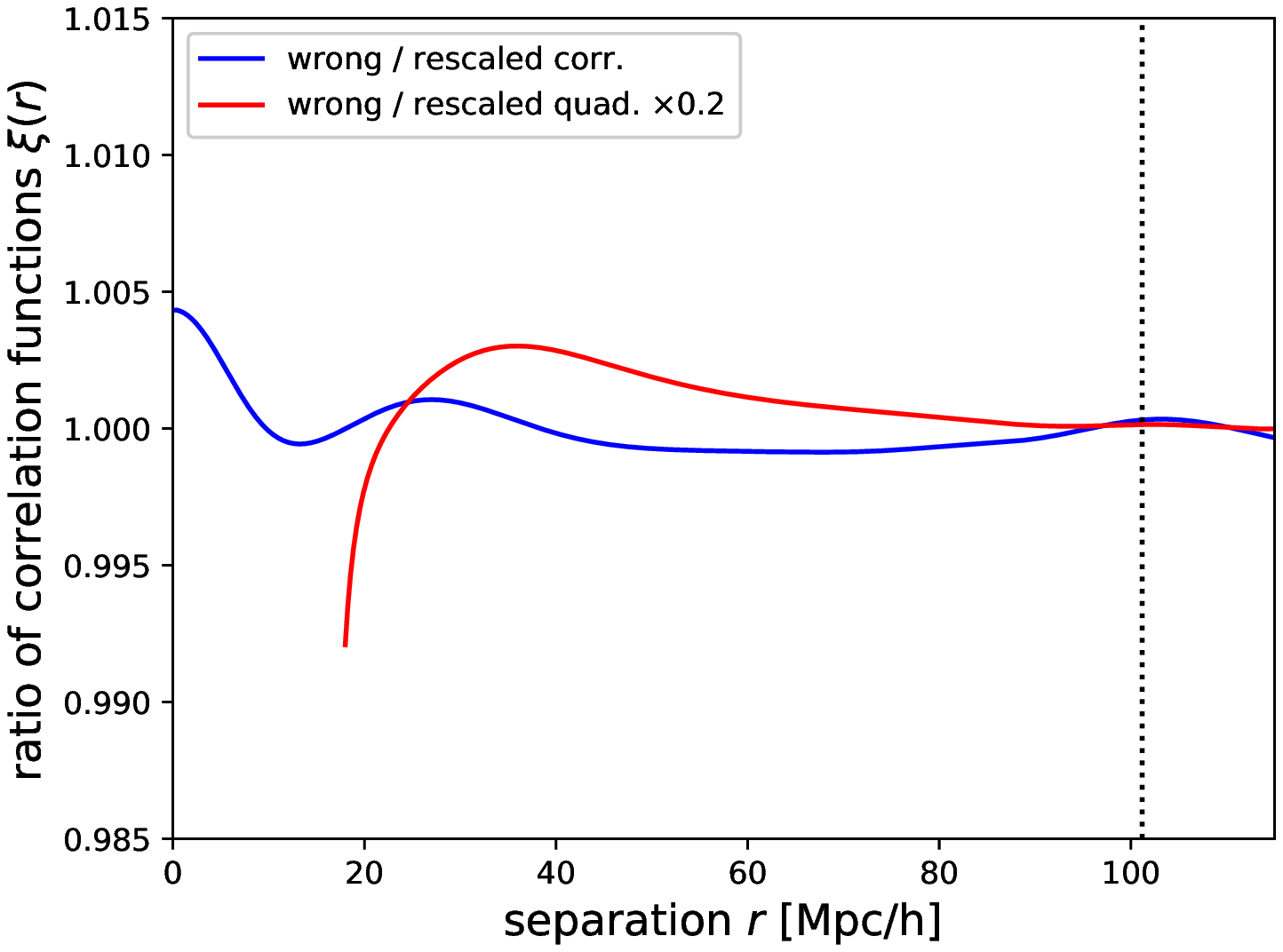}}\\
     \subfigure[Wrong $b=1.1\times 2$: ratio
     ]{\includegraphics[width=.48\textwidth,clip]{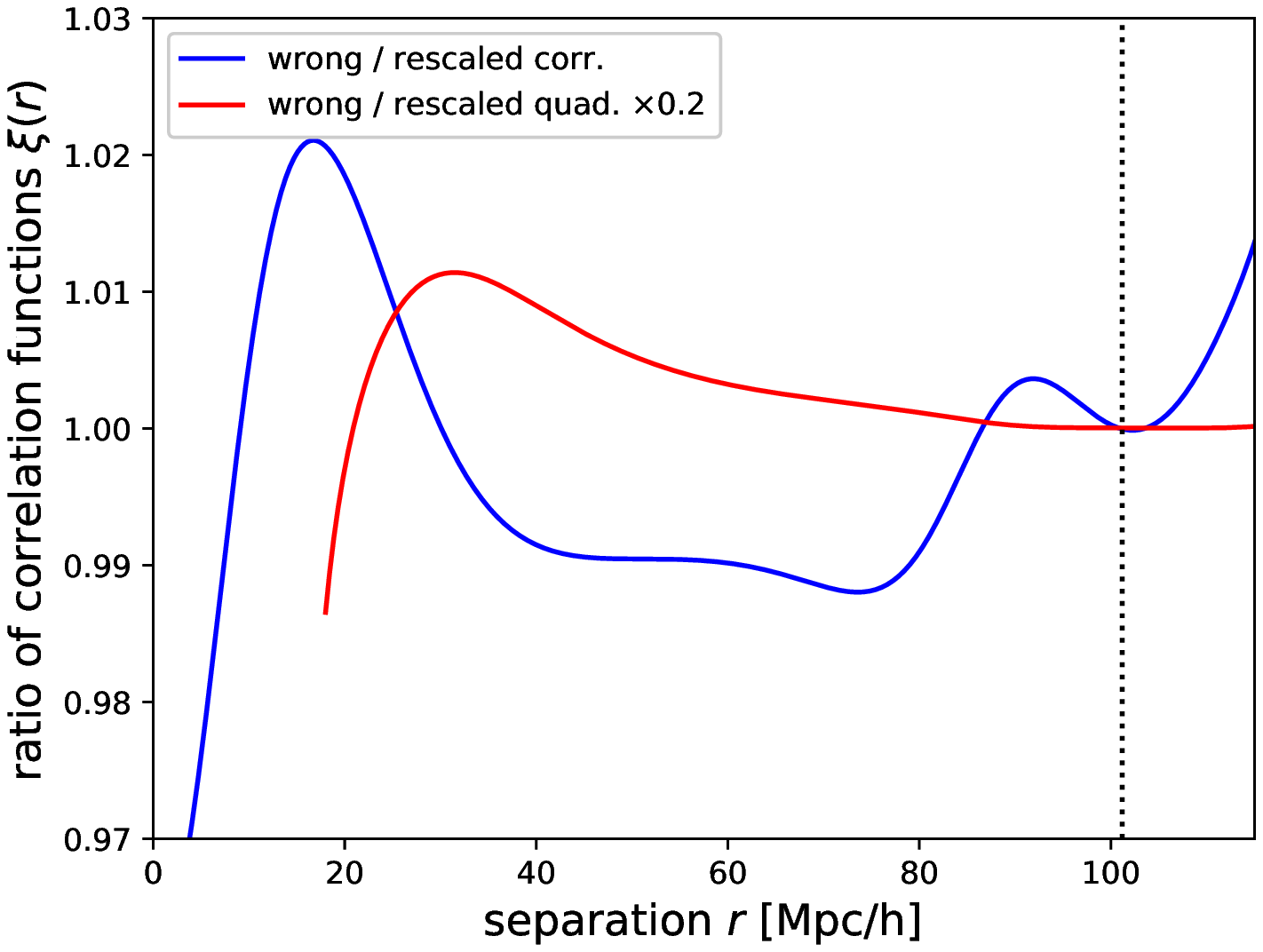}}
     \subfigure[Wrong $f=1.3\times 0.55$: ratio
     ]{\includegraphics[width=.48\textwidth,clip]{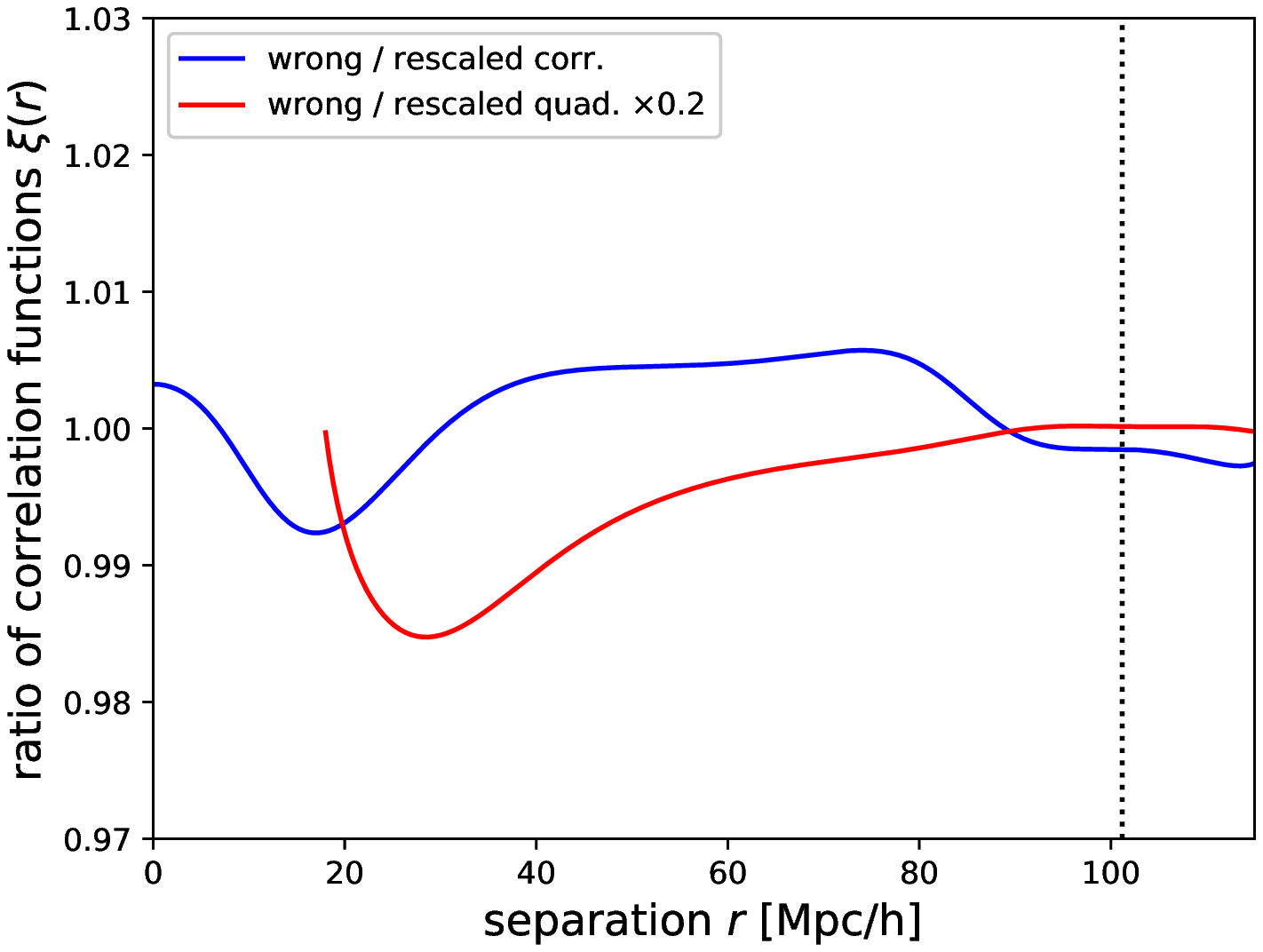}}\\
     \label{tbf} \caption{The ratio of the wrongly reconstructed correlation function to the rescaled correctly reconstructed correlation function. As before, blue indicates the monopole, red the quadrupole. The deviation from unity of the quadrupole has been reduced by a factor $5$ to fit on the same plot scale (it is also only shown for $r>18 \mathrm{Mpc}/h$, since the zero-crossing of the quadrupole at smaller $r$ makes ratios somewhat unstable). It can be seen that no appreciable change in the BAO peak scale results from wrong reconstruction, though the peak shape and quadrupole amplitude can be changed somewhat.}
\label{fig:Xi_all_ratio}
\end{figure*}

To gain further insight, a configuration space picture is helpful. Appropriately Fourier transforming the power spectra, we obtain the equivalent configuration space correlation functions, which are shown in Fig.~\ref{fig:Xi_all}. We also show the ratio of the correctly and incorrectly reconstructed correlation functions in Fig.~\ref{fig:Xi_all_ratio}. 

Fig.~\ref{fig:Xi_all_ratio} is particularly illuminating, as it confirms and clarifies our previous discussion. It can be seen from the plots of the correlation function ratios (as well as the correlation functions themselves in Fig.~\ref{fig:Xi_all}) that the BAO peak position is only negligibly affected, with the ratio fairly flat close to the BAO peak scale. Quantifying this, we find that the peak position is only affected at the negligible level of less than $\sim 2 \times 10^{-4}$, consistent with our Fourier space calculation. However, we see that incorrect reconstruction affects the peak and broadband shapes at the percent level.

Why is the shift in the BAO position so small? To gain more understanding, we Taylor expand the analytical expressions for the reconstructed power spectrum (Eq.~\ref{eqn:Pfalse_rec}) in small errors $\mathbf{\Delta}$ in parameters such as distances $\mathbf{A}=\mathbf{1}+\mathbf{\Delta}$. If all first order corrections cancelled, this would provide an explanation for why shifts in the BAO peak scale are so small. However, we find that the linear terms of order $\mathbf{\Delta}$ do not directly cancel when expanding the reconstructed power spectrum; this is consistent with our observation that the broadband is affected substantially by wrong reconstruction (with changes of a few percent, i.e.~the same order as the fractional error in distances or other parameters). However, in the limit where the peaks are very narrow -- much narrower than the scale over which the damping function varies -- we can gain more intuition for the robustness of the position of the BAO peak. One of our key results is that wrong reconstruction only substantially modifies the damping factor $D$ in Eq.~\ref{eqn:Pfalse_rec}, beyond the usual Alcock-Paczynsky rescaling of the biased power spectrum ($P(\bk) \rightarrow P(\mathbf{A}\bk)$). For sharp peaks, simply modifying the slowly-varying damping factor, $D\rightarrow D + \mathbf{\Delta} \cdot \partial D / \partial \mathbf{\Delta}$, cannot significantly shift the positions of any of the peaks, because the location of an unperturbed maximum or minimum cannot be affected by a sufficiently smooth modulation. While the Alcock-Paczynski rescaling does change the peak positions in the biased power spectrum $P$ to leading order in $\mathbf{\Delta}$, in this case the first order effect is removed by comparing the peak positions to those in the rescaled, correctly reconstructed power spectrum. Since the first order effects in both $P$ and $D$ are thus removed, the leading order effect is a second order (or higher) one in $\mathbf{\Delta}$ -- consistent with the observation of $\sim 10^{-4}$ shifts in the peak positions for a $3\%$ error in the distances.

We now turn to the magnitude of the quadrupole. From Fig.~\ref{fig:Xi_all_ratio} (as well as Fig.~\ref{fig:Pk_all_diff}), we note that the size and shape of the quadrupole is affected at the $\sim 5\%$ level by even moderate errors in the fiducial cosmology, RSD parameters and biases. Some care must thus be taken when using quadrupole information, especially at small separations $r$, and incorrect reconstruction assumptions should be considered and investigated as a cause of anomalous quadrupole signals.

We also investigate the impact of using a larger smoothing scale $\Sigma_{\mathrm{sm}}=20~\mathrm{Mpc}/h$ and a smaller scale $10~\mathrm{Mpc}/h$ than that of our default analysis ($\Sigma_{\mathrm{sm}}=14.1 ~\mathrm{Mpc}/h$). Our previous results for the BAO peak shift still hold in both cases -- the shifts we obtain are at the level of $3 \times 10^{-4}$ or less and thus negligible. The detailed form of the correlation function, however, is affected at some level by a different scale choice. In particular, the deviations from unity in the correlation function ratio as shown in Fig.~3 increase somewhat (by a factor 1-2 for the monopole, 2-3 for the quadrupole) when we use the smaller $10~\mathrm{Mpc}/h$ smoothing scale; when using the larger smoothing scale, the departures from unity are reduced by a factor 1-2 for the monopole and $\approx$2 for the quadrupole. Despite this, our results are qualitatively similar in all cases and our conclusions are thus not strongly dependent on the smoothing scale.

A caveat to our analysis is that we have only expanded the  non-zero-lag terms in our derivation to leading order in perturbation theory. At higher order, mode-coupling terms are known to enter and lead to shifts in the BAO peak position. However, as mentioned previously, assuming that the errors in the assumed cosmology, bias or RSD parameters are fairly small, any modification of the mode-coupling part will arise from a product of these small errors and the small higher order mode-coupling terms; such terms are thus expected to be negligible.

How do our results compare to previous work based on simulations? In \cite{Xu13}, the authors find mean shifts in the BAO scale of order $\leq 10^{-3}$ when assuming 20\% errors on bias and RSD parameters, but given the scatter on their results these shifts do not differ from zero with statistical significance. Errors when assuming a significantly wrong cosmology ($\Omega_m = 0.4$) are larger, but the authors use a linearized template that, the authors note, is expected to fail for large deviations in cosmology. Ref.~\cite{Var18} perform a similar analysis with a wrong fiducial cosmology which differs by $0.5\%$ in the matter density $\Omega_m$. They find a shift of $\approx 9\times 10^{-4}$ for wrong reconstruction. Ref.~\cite{Meh15} also investigated BAO reconstruction in presence of errors on the bias parameters, but they focus on the damping and do not report potential shifts.

Thus, while our results are generally in agreement with simulated results, in a few cases the simulations give slightly higher values of the spurious shifts in the BAO scale. One possible explanation for this is that the shape of the BAO peak (or oscillation envelope in Fourier space) is modified in our calculation, which could lead to non-negligible shifts when fitting with an unmodified template. Our work suggests that a straightforward solution to this problem would be to use Eq.~(\ref{eqn:Pfalse_rec}) or a fitting function of a similar form as the expectation value for the power spectrum in constructing a likelihood. We note that the information content might be slightly enhanced by such a choice, as the peak shape or oscillation envelope may contain some information about parameters such as $\alpha_\perp$. However, it is unclear how much of this information would still be present after marginalizing over all nuisance parameters (such as the damping scale).

As an alternative to modifying the likelihood, one could consider marginalizing the power spectrum measurement not just over an additive broadband component, but also a multiplicative component in Fourier space. Other solutions might include iterating the reconstruction until the cosmology is converged; given that the effects of wrong reconstruction are small, we expect convergence to be rapid (although propagating noise fluctuations through reconstruction could potentially also cause small biases). We defer an exploration of all these ideas to future work.

\section{Conclusions}
\label{sec:conclusions}

In this paper we have analytically calculated how wrong assumptions about the fiducial cosmology, bias and RSD parameters impact the reconstructed power spectrum. We have derived a full expression for this power spectrum (Eq.~\ref{eqn:Pfalse_rec}) and have evaluated it for realistic scenarios.

Assuming large but not unrealistic reconstruction errors ($3\%$ on distances, $10\%$ on bias, $30\%$ on RSD) we find that:
\begin{itemize}
\item The shifts in the BAO monopole peak position(s) in both Fourier and configuration space appear negligible. 
\item The shape of the BAO peak or the oscillation envelope in Fourier space can be modified at the percent level. Fitting with a wrong template could potentially lead to small errors in parameter inference; possible solutions include incorporating a wrong-reconstruction template (from Eq.~\ref{eqn:Pfalse_rec}) in the likelihood, marginalizing over an oscillation envelope shape in a Fourier space analysis, or potentially iterating the reconstruction.
\item The quadrupole can be affected at the $5\%$ level by even moderate errors in the assumed cosmology, bias and RSD.
\end{itemize}
Future research directions include investigating new likelihoods or iterative fitting and examining whether wrong reconstruction impacts BAO constraints on $N_\mathrm{eff}$, the effective number of neutrinos \cite{NeffTh,NeffMeas}.

\acknowledgments{We thank Hee-Jong Seo and Florian Beutler for helpful comments on the draft as well as useful discussions. BDS acknowledges support from an Isaac Newton Trust Early Career Grant as well as an STFC Ernest Rutherford Fellowship. BDS also thanks Lawrence Berkeley Laboratory and the Berkeley Center for Cosmological Physics for hospitality during the completion of this work.  MW is supported by the DOE and by NSF.}

\appendix
\section{Appendix: Evaluating the incorrectly reconstructed spectra}

In this appendix, we will describe in detail the full, general calculation of the incorrectly reconstructed power spectrum outlined in section \ref{sec:everything_wrong}. Following our discussion in section \ref{sec:everything_wrong}, we must evaluate three terms ($P_{dd}$, $P_{ds}$ and $P_{ss}$). For illustration we will focus on the $P_{dd}$ term, which is given by
\beqn
\int d^3q \frac{d\lambda_1}{2\pi} \frac{d\lambda_2}{2\pi}  F^*(\lambda_1) F(\lambda_2)  \ e^{-i\bk'\cdot\bq}\left\langle \exp\left[ -i\bk' \cdot \mathbf{Z}_{*}\bP(\bq_2)+i\bk'\cdot \mathbf{Z}_{*}\bP(\bq_1) + i\lambda_2\delta(\bq_2) - i\lambda_1\delta(\bq_1) \right]\right\rangle \nonumber 
\eeqn
where we have defined $\mathbf{Z}\equiv\mathbf{R-\tilde R \tilde S}$. We can evaluate the expectation value with the cumulant theorem as before
\beqn
&& \left\langle \exp\left[ -i\bk'\cdot \mathbf{Z}_{*}\bP(\bq_2) + i\bk'\cdot \mathbf{Z}_{*}\bP(\bq_1) + i\lambda_2\delta(\bq_2) -i\lambda_1 \delta(\bq_1)\right] \right\rangle \nonumber \\
&&= \exp\left[-\frac{1}{2} \left\langle \lambda_1^2 \delta^2(\bq_1) + \lambda_2^2 \delta^2(\bq_2) + 2 \lambda_1 \delta(\bq_1) \bk' \cdot \mathbf{Z}_{*}\bP(\bq_2) \right.\right. \nonumber \\
&&+ \left.\left.  2 \lambda_2 \delta(\bq_2) \bk'\cdot \mathbf{Z}_{*} \bP(\bq_1) - 2 \lambda_1\lambda_2\delta(\bq_1)\delta(\bq_2) \right\rangle \right]\nonumber \\&&\times\exp\left[-\frac{1}{2}\left\langle  \left( \bk' \cdot  \mathbf{Z}_{*}\bP( \bq_1)-\bk' \cdot \mathbf{Z}_{*}\bP(\bq_2) \right)^2 \right\rangle \right]
\eeqn

We will begin by discussing the non-zero-lag terms. Expanding the exponential, we can perform the $q$ and $\lambda$ integrals for all the terms (while keeping the zero-lag parts exponentiated as previously).  For example
\begin{align} 
& \int d^3q\ e^{-i\bk'\cdot\bq}
  \left\langle \delta(\bq_1) \bk'\cdot \mathbf{Z}_{*}\bP(\bq_2) \right\rangle \nonumber \\
&= k'_j \int d^3q \frac{d\bk_1 d\bk_2}{(2\pi)^6}
\ e^{-i\bk'\cdot\bq+i\bk_1\cdot\bq_1 + i\bk_2\cdot\bq_2} \left[\frac{i\mathbf{Z}\bk_2}{k_2^2}\right]_j
\left\langle \delta(\bk_1) \delta(\bk_2) \right\rangle \nonumber \\
&= k'_j \int d^3q\frac{d\bk_1}{(2\pi)^3} e^{-i\bk'\cdot\bq+i \bk_1\cdot\bq} \left[\frac{-i\mathbf{Z}\bk_1}{k_1^2} \right]_j P_L(k_1) \nonumber \\
&= -i\frac{[\bk' \cdot \mathbf{Z}\bk']}{k'^2} P_L(k')
\end{align} 
and similarly for the other terms.  For the non-zero-lag part of $P_{dd}$, in the end we obtain
\beqn
\frac{[\bk'\cdot \mathbf{Z}\bk']^2}{k'^4}P_L(k')
+ 2 \frac{[\bk'\cdot\mathbf{Z}\bk']}{k'^2}\langle F'\rangle P_L(k') + \langle F'\rangle^2 P_L(k').
\eeqn
To simplify factors such as $[\bk'\cdot\mathbf{Z} \bk'] =[\bk'\cdot\mathbf{(R-\tilde R \tilde S)} \bk'] $, we note that
\beq
\bk'\cdot\mathbf{\tilde{R}~A^{-2}} \bk' \frac{ k^{'2}}{|\mathbf{A}^{-1}\bk'|^2} = (\mathbf{A^{-1}}\bk') \cdot\mathbf{\tilde R}(\mathbf{A^{-1}}\bk') \frac{ k^{'2}}{|\mathbf{A}^{-1}\bk'|^2}
= \left(1 + \tilde{f}\mu_{\mathbf{A^{-1}}\bk}^2\right)k^{'2},
\eeq
where we have made use of the fact that $\nu^2\equiv\mu_{\mathbf{A^{-1}}\bk}^2 = \left[(\mathbf{A^{-1}}\bk) \cdot\mathbf{\hat z}/|A^{-1}\bk|\right]^2$ in the second equality.
Since $[\bk'\cdot\mathbf{R} \bk']=(1+f\mu'^2)k^{'2}$, it follows that
\begin{equation}
\frac{[\bk'\cdot\mathbf{Z}\bk']}{k'^2} = \left[(1+f\mu'^2)- (1+\tilde f\nu^2) \left[\frac{b+f\mu'^2}{{\tilde b+\tilde f\nu^2}}\right] S(\mathbf{A^{-1}}\bk')\right].
\end{equation}
Using $\langle F'\rangle = b-1$ we can now evaluate the non-zero-lag part of the $P_{dd}$ term
\begin{align}
&\hphantom{=} \frac{[\bk'\cdot\mathbf{Z}\bk']^2}{k'^4} P_L(k') + 
2 \frac{[\bk'\cdot\mathbf{Z}\bk']}{k'^2}\langle F'\rangle P_L(k') +
\langle F'\rangle^2 P_L(k') \nonumber \\
&= \left(\left[b-1\right] + \frac{[\bk'\cdot\mathbf{Z}\bk']}{k'^2}\right)^2 P_L(k') \nonumber \\
&=
[b+f\mu'^2]^2 \left[1 -  \left[ \frac{1+ \tilde f \nu^2 }{\tilde b+ \tilde f \nu^2 }\right] S(\mathbf{A}^{-1} \bk') \right]^2 P_L(k').
\end{align}
Performing a similar analysis for the $P_{ss}$ term, the non-zero-lag piece is
\beqn
\frac{\left[\bk'\cdot\mathbf{\tilde R~\tilde S} \bk'\right]^2}{k'^4}P_L(k') = [b+f\mu'^2]^2 \left[ \frac{1+\tilde f\nu^2}{\tilde b+\tilde f\nu^2} S(\mathbf{A}^{-1} \bk')\right]^2 P_L(k').\nonumber
\eeqn
For $P_{ds}$, the relevant piece is
\beqn
&& \frac{[\bk'\cdot\mathbf{Z} \bk']\left[\bk'\cdot \mathbf{\tilde R\tilde S} \bk'\right]}{k^4} P_L(k') +  \langle F'\rangle \frac{\left[\bk'\cdot\mathbf{\tilde R\tilde S}\bk'\right]}{k'^2} P_L(k') \nonumber\\
&=& \left( \left[b-1\right] + \frac{[ \bk'\cdot\mathbf{Z}\bk']}{k'^2}\right) \frac{\left[\bk'\cdot\mathbf{\tilde R\tilde S} \bk'\right]}{k'^2}P_L(k')\nonumber \\
&=& [b+f\mu^2]^2 \left[1 - \frac{1+\tilde f\nu^2}{\tilde b+\tilde f\nu^2} S(\mathbf{A}^{-1} \bk')\right] \left[ \frac{1+\tilde f\nu^2}{\tilde b+\tilde f\nu^2} S(\mathbf{A}^{-1} \bk')\right] P_L(k').
\eeqn

Finally, we consider the zero-lag piece of all three terms. For this calculation, it will be convenient to define a new function $g$ as follows:
\beqn
\mathbf{S}(\bk' ) =  \mathbf{A}^{-2}   \left[\frac{ k^{'2}}{|\mathbf{A}^{-1} \bk' |^2}   \frac{ b+ f\mu'^2}{{\tilde b+\tilde f\nu^2}}  \right]  S(\mathbf{A}^{-1} \bk')\equiv \mathbf{A}^{-2}  g(\bk')  S(\mathbf{A}^{-1} \bk'),
\eeqn
where $g(\bk')$ is the factor in square brackets. For the zero-lag piece of the $P_{ss}$ term, we obtain
\beqn 
&=&\exp \left[- \sum_{i} k_i^{'2} \int   \frac{d \bk_1}{(2\pi)^3} \left[ \mathbf{\tilde R A^{-2}} \bk_1 \right]_i^2  g^2(\bk_1)S^2(\mathbf{A^{-1}}\bk_1) \frac{P_L(k_1)}{k_1^4} \right]\nonumber \\
&=&\exp \bigg[-k_\bot^{'2} \int\frac{d \bk_1}{(2\pi)^3} (1-\mu_1^2)k_{1}^2  g^2(\bk_1)S^2(\mathbf{A^{-1}} \bk_1) \frac{P_L(k_1)}{2 k_1^4 \alpha_\bot^4 } \nonumber \\
&-& k_\parallel^{'2}(1+\tilde f)^2 \int\frac{d\bk_1}{(2\pi)^3} \mu_1^2 k_{1}^2  g^2(\bk_1)S^2(\mathbf{A^{-1}} \bk_1) \frac{P_L(k_1)}{k_1^4 \alpha_\parallel^4 } \bigg] \nonumber\\&
&\equiv \exp \left[-\frac{1}{2}k_\bot^{'2} \Sigma^{2~\tilde f, \tilde b, \mathbf{A}}_{ss,\bot} - \frac{1}{2}k_\parallel^{'2} (1+\tilde f)^2 \Sigma^{2~\tilde f, \tilde b, \mathbf{A}}_{ss,\parallel}\right],
\label{defS1}
\eeqn
where the last line defines new damping functions from the two terms in the previous exponent. Similarly, for $P_{dd}$ we obtain:
\beqn 
&& \exp \bigg[-k_\bot^{'2} \int   \frac{d \bk_1}{(2\pi)^3}  (1-\mu^2) k_{1}^2 (1- \frac{g(\bk_1)S(\mathbf{A^{-1}}\bk_1)}{\alpha_\bot^2})^2\frac{P_L(k_1)}{2 k_1^4 } \nonumber \\
&-& k_\parallel^{'2} (1+f)^2 \int\frac{d \bk_1}{(2\pi)^3} \mu^2 k_{1}^2 (1- \frac{1+\tilde f}{(1+ f)\alpha_\parallel^2}g(\bk_1)S(\mathbf{A^{-1}}\bk_1))^2 \frac{P_L(k_1)}{k_1^4  } \bigg] \nonumber \\
&\equiv & \exp \left[-\frac{1}{2}k_\bot^{'2} \Sigma^{2~\tilde f, \tilde b, \mathbf{A}}_{dd,\bot} - \frac{1}{2} k_\parallel^{'2} (1+f)^2 \Sigma^{2~\tilde f, \tilde b, \mathbf{A}}_{dd,\parallel}\right]
\label{defS2}
\eeqn
The exponent of the $P_{ds}$ term is given by the average of the $P_{ss}$ and $P_{dd}$ term exponents as previously.

Assembling these results gives our most general equation for the reconstructed power spectrum with errors in angle, distance, bias, and RSD. The final expression is:
\begin{align}
P_{\mathrm{rec}}^f(\bk) &= \alpha_\bot^2 \alpha_\parallel \left[b+f\mu^{'2}\right]^2 P_L(k') \left\{
\left[ \frac{1+\tilde f\nu^2}{\tilde b+\tilde f\nu^2} S(k)\right]^2 \mathcal{D}_{ss} \right. \nonumber \\
&+ \left. \left[1-  \frac{1+ \tilde f \nu^2 }{\tilde b+ \tilde f \nu^2 } S(k) \right]^2 \mathcal{D}_{dd}
+  2\left[1 - \frac{1+\tilde f\nu^2}{\tilde b+\tilde f\nu^2} S(k)\right] \left[ \frac{1+\tilde f\nu^2}{\tilde b+\tilde f\nu^2} S(k)\right]  \mathcal{D}_{ds} \right\},
\label{eqn:Pfalse_recA}
\end{align}
where 
\begin{align}
-2\ln\mathcal{D}_{ss} &= k_\bot^{'2} \Sigma^{{2~\tilde f, \tilde b, \mathbf{A}}}_{ss,\bot} + k_\parallel^{'2} (1+\tilde f)^2 \Sigma^{{2~\tilde f, \tilde b, \mathbf{A}}}_{ss,\parallel} \\
-2\ln\mathcal{D}_{dd} &= k_\bot^{'2} \Sigma^{{2~\tilde f, \tilde b, \mathbf{A}}}_{dd,\bot} + k_\parallel^{'2} (1+f)^2 \Sigma^{{2~\tilde f, \tilde b, \mathbf{A}}}_{dd,\parallel} \\
-2\ln\mathcal{D}_{ds} &= \frac{1}{2}\left[
-2\ln\mathcal{D}_{dd} - 2\ln\mathcal{D}_{ss} \right],
\end{align}
the damping coefficients $\Sigma$ are defined in Eqs. (\ref{defS1}-\ref{defS2}) above, and we remind the reader that $\bk =\mathbf{A^{-1}k'}$ and $\nu=\hat{\bk}\cdot \hat{\mathbf{z}}$.

\bibliographystyle{JHEP}
\bibliography{main}

\providecommand{\href}[2]{#2}\begingroup\raggedright\begin{thebibliography}{10}

\bibitem{BOSS_DR12}
S.~{Alam}, M.~{Ata}, S.~{Bailey}, F.~{Beutler}, D.~{Bizyaev}, J.~A. {Blazek},
  A.~S. {Bolton}, J.~R. {Brownstein}, A.~{Burden}, C.-H. {Chuang},
  J.~{Comparat}, A.~J. {Cuesta}, K.~S. {Dawson}, D.~J. {Eisenstein},
  S.~{Escoffier}, H.~{Gil-Mar{\'{\i}}n}, J.~N. {Grieb}, N.~{Hand}, S.~{Ho},
  K.~{Kinemuchi}, D.~{Kirkby}, F.~{Kitaura}, E.~{Malanushenko},
  V.~{Malanushenko}, C.~{Maraston}, C.~K. {McBride}, R.~C. {Nichol}, M.~D.
  {Olmstead}, D.~{Oravetz}, N.~{Padmanabhan}, N.~{Palanque-Delabrouille},
  K.~{Pan}, M.~{Pellejero-Ibanez}, W.~J. {Percival}, P.~{Petitjean},
  F.~{Prada}, A.~M. {Price-Whelan}, B.~A. {Reid}, S.~A.
  {Rodr{\'{\i}}guez-Torres}, N.~A. {Roe}, A.~J. {Ross}, N.~P. {Ross},
  G.~{Rossi}, J.~A. {Rubi{\~n}o-Mart{\'{\i}}n}, A.~G. {S{\'a}nchez},
  S.~{Saito}, S.~{Salazar-Albornoz}, L.~{Samushia}, S.~{Satpathy}, C.~G.
  {Sc{\'o}ccola}, D.~J. {Schlegel}, D.~P. {Schneider}, H.-J. {Seo},
  A.~{Simmons}, A.~{Slosar}, M.~A. {Strauss}, M.~E.~C. {Swanson}, D.~{Thomas},
  J.~L. {Tinker}, R.~{Tojeiro}, M.~{Vargas Maga{\~n}a}, J.~A. {Vazquez},
  L.~{Verde}, D.~A. {Wake}, Y.~{Wang}, D.~H. {Weinberg}, M.~{White}, W.~M.
  {Wood-Vasey}, C.~{Y{\`e}che}, I.~{Zehavi}, Z.~{Zhai}, and G.-B. {Zhao}, {\it
  {The clustering of galaxies in the completed SDSS-III Baryon Oscillation
  Spectroscopic Survey: cosmological analysis of the DR12 galaxy sample}},
  {\em ArXiv e-prints} (July, 2016)
  [\href{http://xxx.lanl.gov/abs/1607.03155}{{\tt arXiv:1607.03155}}].

\bibitem{Wei13}
D.~H. {Weinberg}, M.~J. {Mortonson}, D.~J. {Eisenstein}, C.~{Hirata}, A.~G.
  {Riess}, and E.~{Rozo}, {\it {Observational probes of cosmic acceleration}},
  {\em \physrep} {\bf 530} (Sept., 2013) 87--255,
  [\href{http://xxx.lanl.gov/abs/1201.2434}{{\tt arXiv:1201.2434}}].

\bibitem{PDG14}
{\bf Particle Data Group} Collaboration, K.~Olive et~al., {\it {Review of
  Particle Physics}},  {\em Chin.Phys.} {\bf C38} (2014) 090001.

\bibitem{ESSS07}
D.~J. {Eisenstein}, H.-J. {Seo}, E.~{Sirko}, and D.~N. {Spergel}, {\it
  {Improving Cosmological Distance Measurements by Reconstruction of the Baryon
  Acoustic Peak}},  {\em \apj} {\bf 664} (Aug., 2007) 675--679,
  [\href{http://xxx.lanl.gov/abs/astro-ph/0604362}{{\tt astro-ph/0604362}}].

\bibitem{PWC09}
N.~{Padmanabhan}, M.~{White}, and J.~D. {Cohn}, {\it {Reconstructing baryon
  oscillations: A Lagrangian theory perspective}},  {\em \prd} {\bf 79} (Mar.,
  2009) 063523, [\href{http://xxx.lanl.gov/abs/0812.2905}{{\tt
  arXiv:0812.2905}}].

\bibitem{NWP09}
Y.~{Noh}, M.~{White}, and N.~{Padmanabhan}, {\it {Reconstructing baryon
  oscillations}},  {\em \prd} {\bf 80} (Dec., 2009) 123501,
  [\href{http://xxx.lanl.gov/abs/0909.1802}{{\tt arXiv:0909.1802}}].

\bibitem{Seo10}
H.-J. {Seo}, J.~{Eckel}, D.~J. {Eisenstein}, K.~{Mehta}, M.~{Metchnik},
  N.~{Padmanabhan}, P.~{Pinto}, R.~{Takahashi}, M.~{White}, and X.~{Xu}, {\it
  {High-precision Predictions for the Acoustic Scale in the Nonlinear Regime}},
   {\em \apj} {\bf 720} (Sept., 2010) 1650--1667,
  [\href{http://xxx.lanl.gov/abs/0910.5005}{{\tt arXiv:0910.5005}}].

\bibitem{Pad12}
N.~{Padmanabhan}, X.~{Xu}, D.~J. {Eisenstein}, R.~{Scalzo}, A.~J. {Cuesta},
  K.~T. {Mehta}, and E.~{Kazin}, {\it {A 2 per cent distance to z = 0.35 by
  reconstructing baryon acoustic oscillations - I. Methods and application to
  the Sloan Digital Sky Survey}},  {\em \mnras} {\bf 427} (Dec., 2012)
  2132--2145, [\href{http://xxx.lanl.gov/abs/1202.0090}{{\tt
  arXiv:1202.0090}}].

\bibitem{TasZal12b}
S.~{Tassev} and M.~{Zaldarriaga}, {\it {Towards an optimal reconstruction of
  baryon oscillations}},  {\em \jcap} {\bf 10} (Oct., 2012) 006,
  [\href{http://xxx.lanl.gov/abs/1203.6066}{{\tt arXiv:1203.6066}}].

\bibitem{McCSza12}
N.~{McCullagh} and A.~S. {Szalay}, {\it {Nonlinear Behavior of Baryon Acoustic
  Oscillations from the Zel'dovich Approximation Using a Non-Fourier
  Perturbation Approach}},  {\em \apj} {\bf 752} (June, 2012) 21,
  [\href{http://xxx.lanl.gov/abs/1202.1306}{{\tt arXiv:1202.1306}}].

\bibitem{SheZal12}
B.~D. {Sherwin} and M.~{Zaldarriaga}, {\it {Shift of the baryon acoustic
  oscillation scale: A simple physical picture}},  {\em \prd} {\bf 85} (May,
  2012) 103523, [\href{http://xxx.lanl.gov/abs/1202.3998}{{\tt
  arXiv:1202.3998}}].

\bibitem{Xu13}
X.~{Xu}, A.~J. {Cuesta}, N.~{Padmanabhan}, D.~J. {Eisenstein}, and C.~K.
  {McBride}, {\it {Measuring D$_{A}$ and H at z=0.35 from the SDSS DR7 LRGs
  using baryon acoustic oscillations}},  {\em \mnras} {\bf 431} (May, 2013)
  2834--2860, [\href{http://xxx.lanl.gov/abs/1206.6732}{{\tt
  arXiv:1206.6732}}].

\bibitem{Sch15}
M.~{Schmittfull}, Y.~{Feng}, F.~{Beutler}, B.~{Sherwin}, and M.~Y. {Chu}, {\it
  {Eulerian BAO reconstructions and N -point statistics}},  {\em \prd} {\bf 92}
  (Dec., 2015) 123522, [\href{http://xxx.lanl.gov/abs/1508.06972}{{\tt
  arXiv:1508.06972}}].

\bibitem{BPH15}
A.~{Burden}, W.~J. {Percival}, and C.~{Howlett}, {\it {Reconstruction in
  Fourier space}},  {\em \mnras} {\bf 453} (Oct., 2015) 456--468,
  [\href{http://xxx.lanl.gov/abs/1504.02591}{{\tt arXiv:1504.02591}}].

\bibitem{AchBla15}
I.~{Achitouv} and C.~{Blake}, {\it {Improving reconstruction of the baryon
  acoustic peak: The effect of local environment}},  {\em \prd} {\bf 92} (Oct.,
  2015) 083523, [\href{http://xxx.lanl.gov/abs/1507.03584}{{\tt
  arXiv:1507.03584}}].

\bibitem{Whi15a}
M.~{White}, {\it {Reconstruction within the Zeldovich approximation}},  {\em
  \mnras} {\bf 450} (July, 2015) 3822--3828,
  [\href{http://xxx.lanl.gov/abs/1504.03677}{{\tt arXiv:1504.03677}}].

\bibitem{Coh16}
J.~D. {Cohn}, M.~{White}, T.-C. {Chang}, G.~{Holder}, N.~{Padmanabhan}, and
  O.~{Dor{\'e}}, {\it {Combining galaxy and 21-cm surveys}},  {\em \mnras} {\bf
  457} (Apr., 2016) 2068--2077, [\href{http://xxx.lanl.gov/abs/1511.07377}{{\tt
  arXiv:1511.07377}}].

\bibitem{Seo16}
H.-J. {Seo}, F.~{Beutler}, A.~J. {Ross}, and S.~{Saito}, {\it {Modeling the
  reconstructed BAO in Fourier space}},  {\em \mnras} {\bf 460} (Aug., 2016)
  2453--2471, [\href{http://xxx.lanl.gov/abs/1511.00663}{{\tt
  arXiv:1511.00663}}].

\bibitem{Var16}
M.~{Vargas-Maga{\~n}a}, S.~{Ho}, A.~J. {Cuesta}, R.~{O'Connell}, A.~J. {Ross},
  D.~J. {Eisenstein}, W.~J. {Percival}, J.~N. {Grieb}, A.~G. {S{\'a}nchez},
  J.~L. {Tinker}, R.~{Tojeiro}, F.~{Beutler}, C.-H. {Chuang}, F.-S. {Kitaura},
  F.~{Prada}, S.~A. {Rodr{\'{\i}}guez-Torres}, G.~{Rossi}, H.-J. {Seo}, J.~R.
  {Brownstein}, M.~{Olmstead}, and D.~{Thomas}, {\it {The clustering of
  galaxies in the completed SDSS-III Baryon Oscillation Spectroscopic Survey:
  theoretical systematics and Baryon Acoustic Oscillations in the galaxy
  correlation function}},  {\em ArXiv e-prints} (Oct., 2016)
  [\href{http://xxx.lanl.gov/abs/1610.03506}{{\tt arXiv:1610.03506}}].

\bibitem{Hik17}
C.~{Hikage}, K.~{Koyama}, and A.~{Heavens}, {\it {Perturbation Theory for BAO
  reconstructed fields: one-loop results in real-space matter density field}},
  {\em ArXiv e-prints} (Mar., 2017)
  [\href{http://xxx.lanl.gov/abs/1703.07878}{{\tt arXiv:1703.07878}}].

\bibitem{Zel70}
Y.~B. {Zel'dovich}, {\it {Gravitational instability: An approximate theory for
  large density perturbations.}},  {\em \aap} {\bf 5} (Mar., 1970) 84--89.

\bibitem{Mat08a}
T.~{Matsubara}, {\it {Resumming cosmological perturbations via the Lagrangian
  picture: One-loop results in real space and in redshift space}},  {\em \prd}
  {\bf 77} (Mar., 2008) 063530, [\href{http://xxx.lanl.gov/abs/0711.2521}{{\tt
  arXiv:0711.2521}}].

\bibitem{Mat08b}
T.~{Matsubara}, {\it {Nonlinear perturbation theory with halo bias and
  redshift-space distortions via the Lagrangian picture}},  {\em \prd} {\bf 78}
  (Oct., 2008) 083519, [\href{http://xxx.lanl.gov/abs/0807.1733}{{\tt
  arXiv:0807.1733}}].

\bibitem{TasZal12a}
S.~{Tassev} and M.~{Zaldarriaga}, {\it {The mildly non-linear regime of
  structure formation}},  {\em \jcap} {\bf 4} (Apr., 2012) 013,
  [\href{http://xxx.lanl.gov/abs/1109.4939}{{\tt arXiv:1109.4939}}].

\bibitem{CLPT}
J.~{Carlson}, B.~{Reid}, and M.~{White}, {\it {Convolution Lagrangian
  perturbation theory for biased tracers}},  {\em \mnras} {\bf 429} (Feb.,
  2013) 1674--1685, [\href{http://xxx.lanl.gov/abs/1209.0780}{{\tt
  arXiv:1209.0780}}].

\bibitem{Bha96}
S.~{Bharadwaj}, {\it {The Evolution of Correlation Functions in the Zeldovich
  Approximation and Its Implications for the Validity of Perturbation Theory}},
   {\em \apj} {\bf 472} (Nov., 1996) 1--+,
  [\href{http://xxx.lanl.gov/abs/astro-ph/9}{{\tt astro-ph/9}}].

\bibitem{CasWhi18}
E.~{Castorina} and M.~{White}, {\it {The Zeldovich approximation and wide-angle
  redshift-space distortions}},  {\em \mnras} {\bf 479} (Sept., 2018) 741--752,
  [\href{http://xxx.lanl.gov/abs/1803.08185}{{\tt arXiv:1803.08185}}].

\bibitem{EisHu}
D.~J. {Eisenstein} and W.~{Hu}, {\it {Power Spectra for Cold Dark Matter and
  Its Variants}},  {\em \apj} {\bf 511} (Jan., 1999) 5--15,
  [\href{http://xxx.lanl.gov/abs/astro-ph/9710252}{{\tt astro-ph/9710252}}].

\bibitem{PlanckParams}
{Planck Collaboration}, P.~A.~R. {Ade}, N.~{Aghanim}, M.~{Arnaud},
  M.~{Ashdown}, J.~{Aumont}, C.~{Baccigalupi}, A.~J. {Banday}, R.~B.
  {Barreiro}, J.~G. {Bartlett}, and et~al., {\it {Planck 2015 results. XIII.
  Cosmological parameters}},  {\em \aap} {\bf 594} (Sept., 2016) A13,
  [\href{http://xxx.lanl.gov/abs/1502.01589}{{\tt arXiv:1502.01589}}].

\bibitem{Var18}
M.~{Vargas-Maga{\~n}a}, S.~{Ho}, A.~J. {Cuesta}, R.~{O'Connell}, A.~J. {Ross},
  D.~J. {Eisenstein}, W.~J. {Percival}, J.~N. {Grieb}, A.~G. {S{\'a}nchez},
  J.~L. {Tinker}, R.~{Tojeiro}, F.~{Beutler}, C.-H. {Chuang}, F.-S. {Kitaura},
  F.~{Prada}, S.~A. {Rodr{\'{\i}}guez-Torres}, G.~{Rossi}, H.-J. {Seo}, J.~R.
  {Brownstein}, M.~{Olmstead}, and D.~{Thomas}, {\it {The clustering of
  galaxies in the completed SDSS-III Baryon Oscillation Spectroscopic Survey:
  theoretical systematics and Baryon Acoustic Oscillations in the galaxy
  correlation function}},  {\em \mnras} {\bf 477} (June, 2018) 1153--1188.

\bibitem{Meh15}
K.~T. {Mehta}, H.-J. {Seo}, J.~{Eckel}, D.~J. {Eisenstein}, M.~{Metchnik},
  P.~{Pinto}, and X.~{Xu}, {\it {Galaxy Bias and Its Effects on the Baryon
  Acoustic Oscillation Measurements}},  {\em \apj} {\bf 734} (June, 2011) 94,
  [\href{http://xxx.lanl.gov/abs/1104.1178}{{\tt arXiv:1104.1178}}].

\bibitem{NeffTh}
D.~{Baumann}, D.~{Green}, and B.~{Wallisch}, {\it {Searching for Light Relics
  with Large-Scale Structure}},  {\em ArXiv e-prints} (Dec., 2017)
  [\href{http://xxx.lanl.gov/abs/1712.08067}{{\tt arXiv:1712.08067}}].

\bibitem{NeffMeas}
D.~{Baumann}, F.~{Beutler}, R.~{Flauger}, D.~{Green}, M.~{Vargas-Maga{\~n}a},
  A.~{Slosar}, B.~{Wallisch}, and C.~{Y{\`e}che}, {\it {First Measurement of
  Neutrinos in the BAO Spectrum}},  {\em ArXiv e-prints} (Mar., 2018)
  [\href{http://xxx.lanl.gov/abs/1803.10741}{{\tt arXiv:1803.10741}}].

\end{thebibliography}\endgroup

\end{document}